\documentclass[aps,prb,preprint,bibliography]{revtex4-1}
\usepackage[utf8]{inputenc}
\usepackage{amsmath}
\DeclareMathOperator{\Tr}{Tr}
\usepackage{physics}
\usepackage{mathtools}
\usepackage{amsfonts}
\usepackage{amssymb}
\usepackage{graphicx}
\usepackage{subcaption}
\usepackage{textcomp}
\usepackage{color}
\usepackage[dvipsnames]{xcolor}
\newcommand{\bk}{\textbf{k}}

\begin{document}

\title{Floquet Engineering of Multifold Fermions}

\author{Rimika Jaiswal}
\affiliation{Undergraduate Programme, Indian Institute of Science, Bangalore 560012, India}
\author{Awadhesh Narayan}
\email{awadhesh@iisc.ac.in}
\affiliation{Solid State and Structural Chemistry Unit, Indian Institute of Science, Bangalore 560012, India}

\date{\today}

\begin{abstract}
Using Floquet theory, we investigate the effect of light on triple fold fermions. We study a low energy model as well as a simplified tight binding model under illumination. 
We find that the three fold degeneracy remains symmetry protected even after applying light if the original band structure is rotationally symmetric around the degeneracy. Otherwise, light can lift the degeneracy and open up a gap. We further investigate the effect of light on the topological Fermi arcs by means of numerical computations. The changes caused by illumination are reflected in experimentally detectable signatures, such as the anomalous Hall conductivity, which we calculate.

\end{abstract}

\maketitle

\section{Introduction}

In the last decades, topological aspects of physics have been in the limelight. Starting with the discovery of the quantum Hall effect~\cite{QHE_Exp_PhysRevLett.45.494,QHE_PhysRevLett.49.405}, ideas from topology have started to permeate condensed matter and materials physics. More recently, the proposal and subsequent discovery of topological insulators and superconductors has led to intense activity~\cite{TI_RevModPhys.82.3045,TIS_RevModPhys.83.1057}. Building on these ideas, the concept of topological semimetals was proposed. They are a class of materials which harbor topologically stable energy band crossing near the Fermi surface. It was found that in materials such as graphene, the low energy behaviour around such degeneracies resembles the relativistic Dirac equation. Soon, condensed matter systems became a basis for the discovery and realisation of fermionic particles predicted in high energy physics such as Majorana fermions, and Dirac and Weyl fermions~\cite{Burkov2016,3dRev_RevModPhys.90.015001}. 

Free fermionic excitations in topological semimetals with no high-energy counterparts were recently discovered~\cite{Bradlynaaf5037}. These novel materials, commonly called multifold semimetals, are characterised by higher order (larger than 2) band crossings at the degenerate points. Experiments have confirmed the presence of such higher order degeneracies in materials such as RhSi~\cite{RhSi_PhysRevLett.119.206401}, CoSi~\cite{CoSi_PhysRevLett.122.076402,CoSi_Rao2019}, AlPt~\cite{AlPt_Schroter2019} and the superconducting metal PdSb$_2$~\cite{PdSb2_doi:10.1002/adma.201906046}. These classes of materials exhibit several remarkable properties. Multifold fermions have an enhanced linear response (compared to Weyl fermions) due to their higher topological charge~\cite{LinCon_PhysRevB.99.155145}. They also show a unique quantized photogalvanic effect~\cite{CPGE_PhysRevB.98.155145}, which has been very recently detected in experiments ~\cite{rees2019observation,ni2020giant,ni2020linear}. 

Floquet engineering~\cite{review_moessner}, or the controlling of topological transitions using periodic drives such as light, has recently become an interesting field of study~\cite{RevGiovannini_2019,reviewoka}. Starting from the work of Oka \textit{et al.}, who discovered that applying light on graphene makes it behave like a topological insulator~\cite{oka2009}, various theoretical studies~\cite{Lindner2011,Oka2011,chiral_chiral,TI_TI_PhysRevLett.110.026603,AN_PhysRevB.91.205445} and experiments~\cite{Wang453,McIver2020,L_pez_2020} have illustrated how new topological phases can be generated by the periodic driving of systems with light. Floquet engineering provides a new way of experimentally realising novel topological phases, with excellent tunability. Promising ways of experimentally realising tunable Weyl-semimetals by shining light on Dirac semimetals~\cite{Weyl_Dirac_Hubener2017}, topological insulators~\cite{Weyl_TI_Wang_2014} or nodal line semimetals~\cite{Weyl_NL_PhysRevB.94.041409,Weyl_NL_PhysRevLett.117.087402,Weyl_NLD_PhysRevB.94.121106} have been demonstrated.
More recently, Floquet theory has also been used to propose possible realisations of fractional Chern insulators~\cite{fracCI_PhysRevLett.112.156801} and topological superconductors~\cite{supercond_PhysRevB.95.134508}. Possible applications of Floquet theory to generate new topological phases in bilayer graphene~\cite{blgraphene_PhysRevResearch.1.023031}, photonic systems~\cite{phth_He2019,phexp_PhysRevA.97.031801} and non-Hermitian systems~\cite{banerjee2020controlling} are also being explored.

Motivated by these developments, here, we study the effect of light on triple fold fermions using the Floquet formalism, based on an effective low energy model, as well as a simplified tight binding model. We show that elliptically polarized light can cause a band gap to open up and/or shift the three fold degeneracy depending on the intrinsic symmetries of the system. We map out the complete phase diagram of triple fold fermions under illumination and calculate the resulting anomalous Hall signatures. We also investigate the effect of light on the topological Fermi arcs by means of numerical computations employing a simplified tight-binding model. Our predicted signatures can be experimentally detected and provide new insights into the interaction between light and multifold fermions. We hope to motivate further exploration of the novel topological phases that can arise from the periodic driving of multifold semimetals.

\section{Effective low energy model}

We consider the general low energy Hamiltonian around a three fold degenerate point as derived by Bradlyn \textit{et al}~\cite{Bradlynaaf5037}. Taking the degenerate point, $k_0$, to be at the origin, the Hamiltonian reads:

\begin{equation}
H_{3f}(\bk) = E_0 + \hbar v_f
\begin{bmatrix}
0 & e^{i\phi} k_x & e^{-i\phi} k_y \\
e^{-i\phi} k_x & 0 & e^{i\phi} k_z \\
e^{i\phi} k_y & e^{-i\phi} k_z & 0
\end{bmatrix},
\label{eq:ham0}
\end{equation}

where $E_0$ is an energy offset, $\hbar$ is the reduced Planck's constant, $v_f$ is the effective velocity around $\bk=0$, and $\phi$ is a material dependent parameter. We shall consider the three band touching point (3-BTP) to be at the Fermi energy and thus set $E_0=0$. From here on, we also work in units where $\hbar, v_f$ and $e$ are unity for simplicity. 

\begin{figure}[h!]
\includegraphics[scale=0.55]{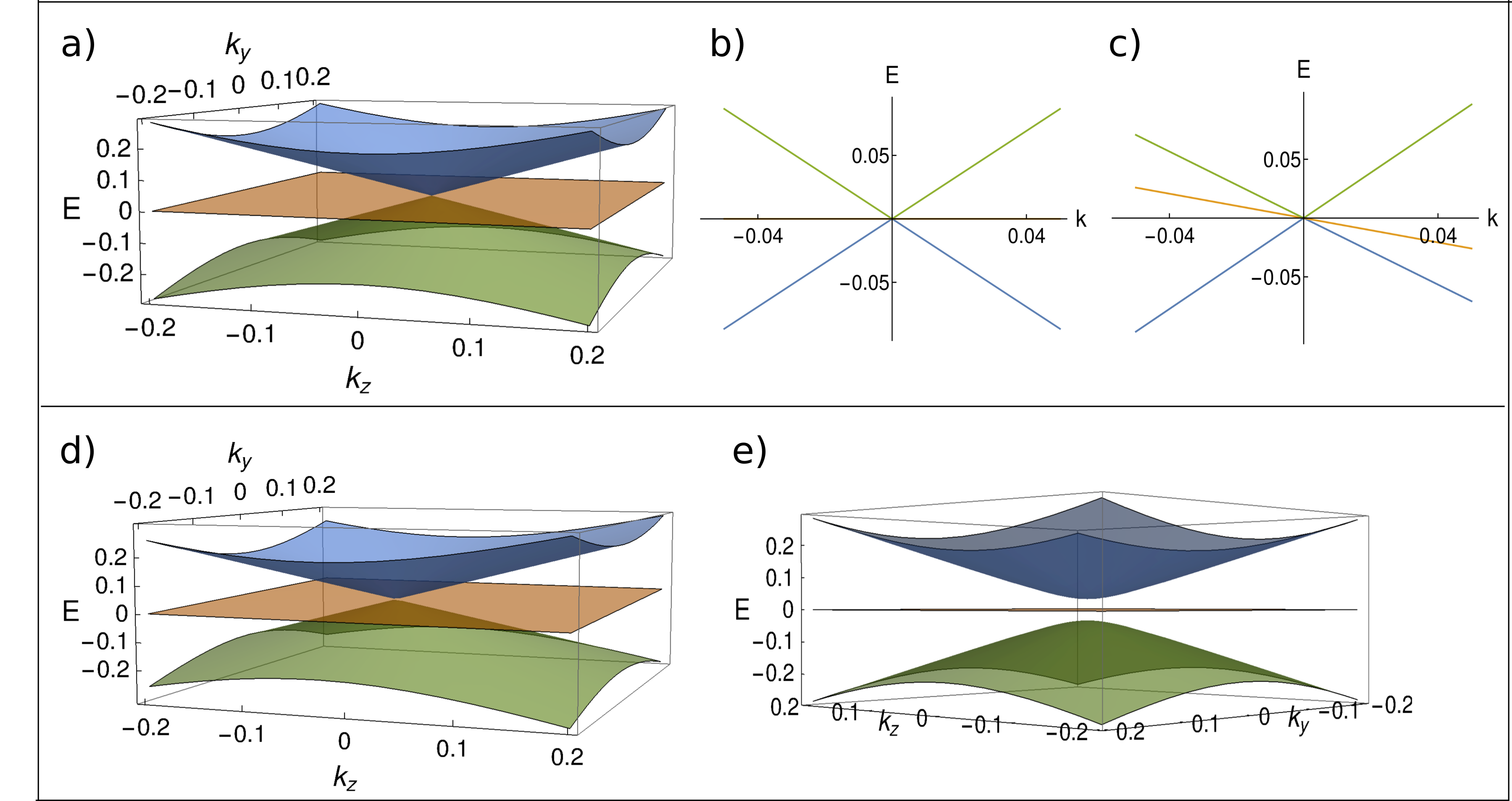}
\caption{Band structure plots illustrating the effect of light on three fold fermions. Top: Band structure before illumination. Bottom: Band structure after illumination. For panels (a), (d) and (e), the eigenvalues are plotted as a function of $k_y$ and $k_z$ (at $k_x=0$). Panel (a) displays the three fold degeneracy at $\bk=0$. Panels (b) and (c) show the dispersion along $k_x=k_y=k_z$ for (b) $\phi=\pi/6$ and (c) $\phi=\pi/12$. The value of $\phi$ determines whether the three fold crossing is centrosymmetric (rotationally symmetric about the degeneracy) or not. Panels (d) and (e) show the resultant band structure after illumination for $\phi=\pi/2$ and $\phi=0$ respectively. For $\phi=\pi/2$, the degeneracy remains intact but shifts along $k_z$ whereas for $\phi=0$ a band gap opens up. Here, we choose the following parameter values: $A_x=2$, $A_y=1$, $\omega=30$, and $\theta=\pi/2$.}
\label{3f bands}
\end{figure}

One can observe that for $\phi=\pi/6$ (mod $\pi/3$) i.e.\ for $\phi=\pi/6,\pi/2,\ldots$, the Hamiltonian takes the form $H=\bk \cdot \pmb S$, where $\pmb S$ is the spin-$1$ representation of SU(2) as given in Equation~\ref{eq:S_def}. The corresponding eigenvalues read $E=0,\pm 
k$, showing that the band-structure is completely centrosymmetric i.e. having complete rotational symmetry about the degeneracy at $\bk=0$. The band structures illustrating this are shown in Figure~\ref{3f bands}.

Away from these $\phi$ values, the conical bands tilt and the centrosymmetry about the degeneracy vanishes. The tilt is maximum for $\phi=0$ (mod $\pi/3$) i.e.\ for $\phi=0, \pi/3,2\pi/3,\ldots$.

The upper and lower bands are topologically non-trivial, having Chern numbers $\pm2$. The flat band in the middle is however trivial with Chern number zero. At half filling, the 3-BTP acts as a source of Berry curvature with charge $+2$ for $\phi\in(0,\frac{\pi}{3})$, whereas, it acts as a sink of Berry curvature with charge $-2$ for $\phi \in (\frac{\pi}{3},\frac{2 \pi}{3})$, and this pattern continues periodically. However, for $\phi=n\frac{\pi}{3},\ n \in \mathbb{Z}$, the eigenvalue spectrum is degenerate along lines corresponding to $|k_x|=|k_y|=|k_z|$ and the Chern number is not defined. These $\phi$ points with extended degeneracy can be thought to partition the space of Hamiltonians into distinct phases that differ in their Fermi surface topology.

\subsection{Floquet analysis for the low energy model}

We next consider shining light on a material with three fold crossing. If a system is subjected to an external time-dependent periodic perturbation such that $H(t+T)=H(t)$, then the Floquet approximation states that in the high frequency ($\omega$) limit, an effective Hamiltonian describing the system can be written as~\cite{review_moessner,reviewoka}

\begin{equation}
H_{floq}(\bk) = H_0(\bk) + \frac{[H_{+1}(\bk),H_{-1}(\bk)]}{
\omega} + \mathcal{O}\left(\frac{1}{ 
\omega^2}\right),
\end{equation}

where $\omega=2\pi/T$ and
$H_{m}(\bk) =\frac{1}{T} \int_{0}^{T} H(\bk,t) e^{-i m \omega t}  dt$. In cases where the Hamiltonian $H(t)$ can be separated into a time-independent and a time-dependent part such that $H(\bk,t)=H_0(\bk)+V(\bk,t)$, the expression simplifies to

\begin{equation}
H_{floq}(\bk) = H_0(\bk) + \frac{[V_{+1}(\bk),V_{-1}(\bk)]}{\omega} + \mathcal{O}\left( \left( V/ \omega\right)^2\right).
\label{eq:floq_formula}
\end{equation}

If light is applied on our three fold fermion system, we can use Floquet theory to study the resulting effective band structure if the frequency of light, $\omega$, is sufficiently large such that $
\omega\gg V$ (see Appendix \ref{section:A_a}). We denote the vector potential of applied light by $\pmb A(t)$. For non-cavity modes, light is plane-polarised, allowing us to choose $A_z=0$. Thus with a phase difference $\theta$ between the $x$ and $y$ components, we have $\pmb A=[A_x \cos(\omega t),A_y \cos(\omega t + \theta),0]$. The effect of this vector potential on our Hamiltonian can be incorporated through the standard Peierls substitution, i.e., $\bk \to \bk+\pmb A$. This gives, $H_{3f}(\bk,t)=H_{3f}(\bk)+V(t)$, where,

\begin{equation}
V(t) = 
\begin{bmatrix}
0 & e^{i\phi}A_x \cos \omega t & e^{-i\phi}A_y \cos(\omega t + \theta)\\
e^{-i\phi}A_x \cos \omega t & 0 & 0 \\
e^{i\phi}A_y \cos(\omega t + \theta) & 0 & 0
\end{bmatrix}.
\end{equation}

Now, for large $\omega$, we can use the Floquet approximation given by Equation \ref{eq:floq_formula} to study the effective photon-dressed static Hamiltonian $H_{floq}$. After integrating over one time period, we obtain the form of the Fourier components as

\begin{equation}
V_{+1}(t) = \frac{ 
1}{2}
\begin{bmatrix}
0 & e^{i\phi} A_x & e^{-i\phi}e^{i\theta} A_y\\
e^{-i\phi} A_x & 0 & 0 \\
e^{i\phi}e^{i\theta}A_y & 0 & 0
\end{bmatrix}
\quad \quad
V_{-1}(t) = \frac{ 
1}{2}
\begin{bmatrix}
0 & e^{i\phi} A_x & e^{-i\phi}e^{-i\theta} A_y\\
e^{-i\phi} A_x & 0 & 0 \\
e^{i\phi}e^{-i\theta}A_y & 0 & 0
\end{bmatrix}.
\end{equation}

This gives the light induced term as

\begin{equation}
\frac{[V_{+1},V_{-1}]}{\hbar \omega}=
\frac{i A_x A_y \sin\theta}{2 \omega}
\begin{bmatrix}
0 & 0 & 0 \\
0 & 0 & -e^{-i2\phi} \\
0 & e^{+i2\phi} & 0
\end{bmatrix}
=i 
\gamma
\begin{bmatrix}
0 & 0 & 0 \\
0 & 0 & -e^{-i2\phi} \\
0 & e^{+i2\phi} & 0
\end{bmatrix},
\end{equation}

where we have defined the prefactor $\gamma$, having the same dimensions as $k$, as

\begin{equation}
\gamma=
\frac{A_x A_y \sin\theta}{2\omega}
\end{equation}

Note that because of the $\sin\theta$ factor, $\gamma$ and hence the light-induced term vanishes for linearly polarized light (i.e. for $\theta=0$), while taking the maximum value for $\theta=\pi/2$ (elliptically polarized light). We can understand this as follows. Elliptical polarization provides the chirality needed to break time-reversal symmetry. In contrast, a linear polarization, made of equal superposition of clockwise and anticlockwise circular polarizations, cannot break time-reversal symmetry and hence does not lead to observable changes.

\subsection{Changes in the band structure}

Our effective Floquet Hamiltonian after applying light thus reads

\begin{equation}
H_{floq}=
\begin{bmatrix}
0 & e^{i\phi} k_x & e^{-i\phi} k_y \\
e^{-i\phi} k_x & 0 & e^{i\phi} k_z-i\gamma e^{-2i\phi} \\
e^{i\phi} k_y & e^{-i\phi} k_z +i\gamma e^{2i\phi} & 0
\end{bmatrix}.
\end{equation}

For certain cases this Hamiltonian can be mapped back to the original low energy model, $H_{3f}$ for three fold fermions. For $\phi=\frac{\pi}{6}$ (mod $\pi/3$), we can write $H_{floq}(\bk)=H_{3f}(\tilde{\bk})$, where $\tilde{\bk}=(k_x,k_y,k_z-\gamma \sin 3\phi)$. The system in presence of light is thus exactly like the unperturbed system but with the 3-BTP now shifted along the $k_z$ axis. We now find the new eigenvalues 

\begin{equation}
E=0,\pm 
\sqrt{k_x^2+k_y^2+(k_z-\gamma \sin 3\phi)^2},
\end{equation}

which are degenerate at $\bk=(0,0,\gamma\sin 3\phi)$.
In contrast, however, for $\phi=0$ (mod $\pi/3$), we observe that a band gap opens up at $\bk=0$ (see Figure~\ref{3f bands}).
For other intermediate values of $\phi$, we find that a band gap opens up along with a shift in the 3-BTP. For $\phi \in (0,\frac{\pi}{3})$, it shifts to the right, while for $\phi \in (\frac{\pi}{3}, \frac{2\pi}{3})$, it shifts to the left, and this alternating pattern continues in a periodic manner. 

Next, we find the analytical expressions for the light-induced band gap ($\Delta_g$) and location of the 3-BTP after the shift ($k_z^0$). The eigenvalues of $H_{floq}$ satisfy the following equation

\begin{equation}
E^3-
E (\gamma^2 + k^2 - 2\gamma k_z \sin 3\phi) - 
2
k_x k_y k _z \cos 3\phi = 0.
\end{equation}

The exact form of the eigenvalues for any general $\phi$ can be derived (given in Appendix \ref{section:A_b}), using which we obtain

\begin{equation}
\Delta_g= 2 
\gamma \cos 3\phi 
\quad \text{and} \quad
k_z^0=\gamma \sin 3\phi.
\end{equation}

\begin{figure}[h!]
	\centering
	\includegraphics[scale=0.65]{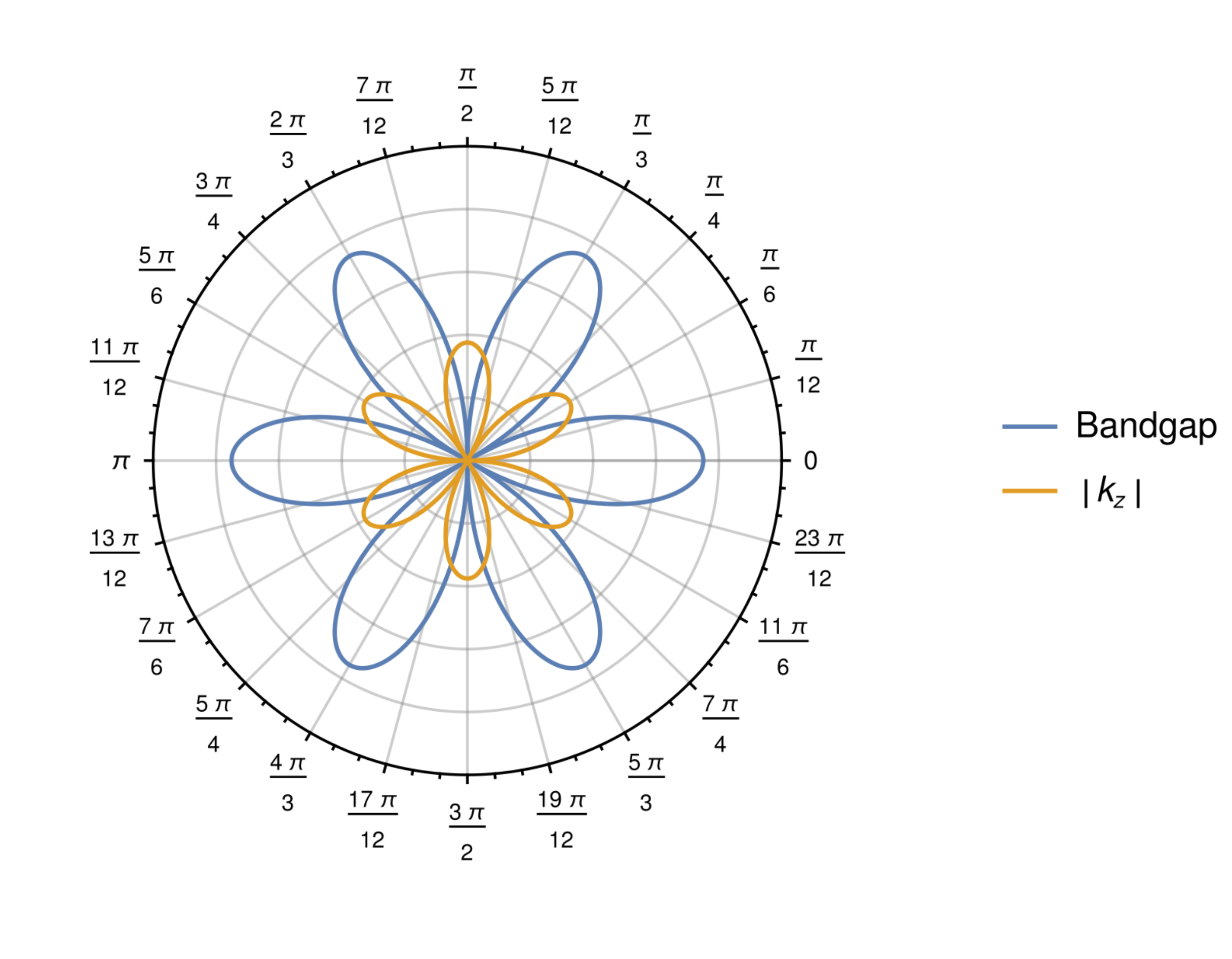}
	\caption{Angular plot depicting the variation of the band-gap ($\Delta_g$) and the shift of the 3 fold degeneracy ($|k_z|$) with $\phi$. There is a $\pi/3$ periodicity in the response of the system. For $\phi=0$ (mod $\pi/3$), a maximum bandgap opens up with the shift being zero. Whereas, for $\phi=\pi/6$ (mod $\pi/3$), there is a maximum shift along $k_z$ but with no bandgap opening. Here we set $\gamma=1/30$}
	\label{variation_with_phi}
\end{figure}

These expressions succinctly summarize the trends described earlier (see Figure \ref{variation_with_phi}). For $\phi=0$ (mod $\pi/3$) we obtain the largest band gap opening $\Delta_g=2\gamma$ at $\bk=0$, while for $\phi=\pi/6$ (mod $\pi/3$), we find the maximum shift of the 3-BTP to $k_z=\gamma$.
The $\cos 3\phi$, $\sin 3\phi$ nature of  variation of the band gap and shift respectively also captures the inherent $\pi/3$ periodicity in the response of the system. We note that the results recently reported in the literature~\cite{3f_pi2_PhysRevB.100.165302} for the special case of $\phi=\pi/2$ are consistent with ours.

Interestingly, we observe the maximum shift of the 3-BTP along the $k_z$-axis, with the three fold degeneracy being maintained, for cases where the original band structure was fully centrosymmetric around the degeneracy. Analogously, we observe the maximum band gap opening up for cases where the original band structure showed the maximal tilt along the $(1,1,1)$ direction. This matches Weyl semimetal studies where it has been shown that lifting the degeneracy in a fully centrosymmetric Weyl semimetal is difficult~\cite{tilt_gap_exp}. It has also been shown that tilted Weyl semimetals have a better response to light, and they can support significant photocurrents while centrosymmetric Weyls cannot~\cite{photocurrent_Weyl_PhysRevB.95.041104}.\\

\textit{Effect of higher order terms:} For our low energy model under illumination, all multi-photon terms $H_m$ for $m\ne 0,1,-1$ are zero. The only non-zero terms in the Floquet-Magnus expansion thus have the form: $[H_0,H_{\pm1}]/\omega$, $[H_{\pm1},[H_0,H_{\mp1}]]/\omega^2$ and so on. All the properties discussed above hold true even when these higher-order terms are considered.

\subsection{Symmetry analysis}
We discovered that depending on the value of $\phi$, the three fold degeneracy can either be lifted or shifted upon illumination with elliptically polarized light. To understand the underlying reason, it is insightful to look at the symmetries of our system, and see if they change after applying light. Multifold degeneracies in space group 199 are characterized by three unitary symmetries which protect the threefold degeneracy: $\{C_{2x}|\bar{\frac{1}{2}} \frac{1}{2} 0 \}$, $\{C_{2y}|0  \frac{1}{2} \bar{\frac{1}{2}} \}$ and $\{C^{-1}_{3,111}|1 0 1 \}$. They can be chosen to be~\cite{Bradlynaaf5037}

\begin{equation}
G_1=
\begin{bmatrix}
-1 & 0 & 0 \\
0 & -1 & 0 \\
0 & 0 & 1
\end{bmatrix},
\quad
G_2=
\begin{bmatrix}
0 & 0 & -1 \\
1 & 0 & 0 \\
0 & -1 & 0
\end{bmatrix},
\quad
G_3=
\begin{bmatrix}
0 & 0 & 1 \\
1 & 0 & 0 \\
0 & 1 & 0
\end{bmatrix}.
\end{equation}

A Hamiltonian obeying these symmetries is constrained to satisfy the following conditions

\begin{equation}
G_i H(k) G_i^{-1} = H[D(G_i)k], \quad
i=1,2,3.
\label{eq:sym_req}
\end{equation}

where, the matrices $D(G_i)$ have the representation

\begin{equation}
D(G_1)=
\begin{bmatrix}
1 & 0 & 0 \\
0 & -1 & 0 \\
0 & 0 & -1 
\end{bmatrix},
\quad
D(G_2)=
\begin{bmatrix}
0 & -1 & 0 \\
0 & 0 & 1 \\
-1 & 0 & 0 
\end{bmatrix},
\quad
D(G_3)=
\begin{bmatrix}
0 & 1 & 0 \\
0 & 0 & 1 \\
1 & 0 & 0
\end{bmatrix}.
\end{equation}

It is easy to verify that our original Hamiltonian $H_{3f}$ satisfies Equation~\ref{eq:sym_req}. Next, we investigate whether these symmetries are preserved once light is applied. We find that Equation~\ref{eq:sym_req} is satisfied by $H_{floq}(\tilde{\bk})$ with $\tilde{\bk}=(k_x,k_y,k_z-\gamma\sin 3\phi)$, whenever $\phi=n \frac{\pi}{6}$ (mod $\pi/3$). This implies that for these values of $\phi$, the three-fold degeneracy shifts but remains symmetry protected even after illumination. On the other hand, for all other values of $\phi$, $H_{floq}(\bk)$ fails to satisfy the symmetry conditions. The degeneracy is no longer symmetry protected and can be lifted by light leading to opening up of a band gap. Note that because both circular and elliptically polarised light break all three lattice symmetries, the results are similar for both these choices of polarisation. Thus, we can understand our results based on these symmetry arguments. 

\subsection{Illustration of band structure engineering with densities of states}

\begin{figure}[h!]
  \centering
  \includegraphics[scale=0.5]{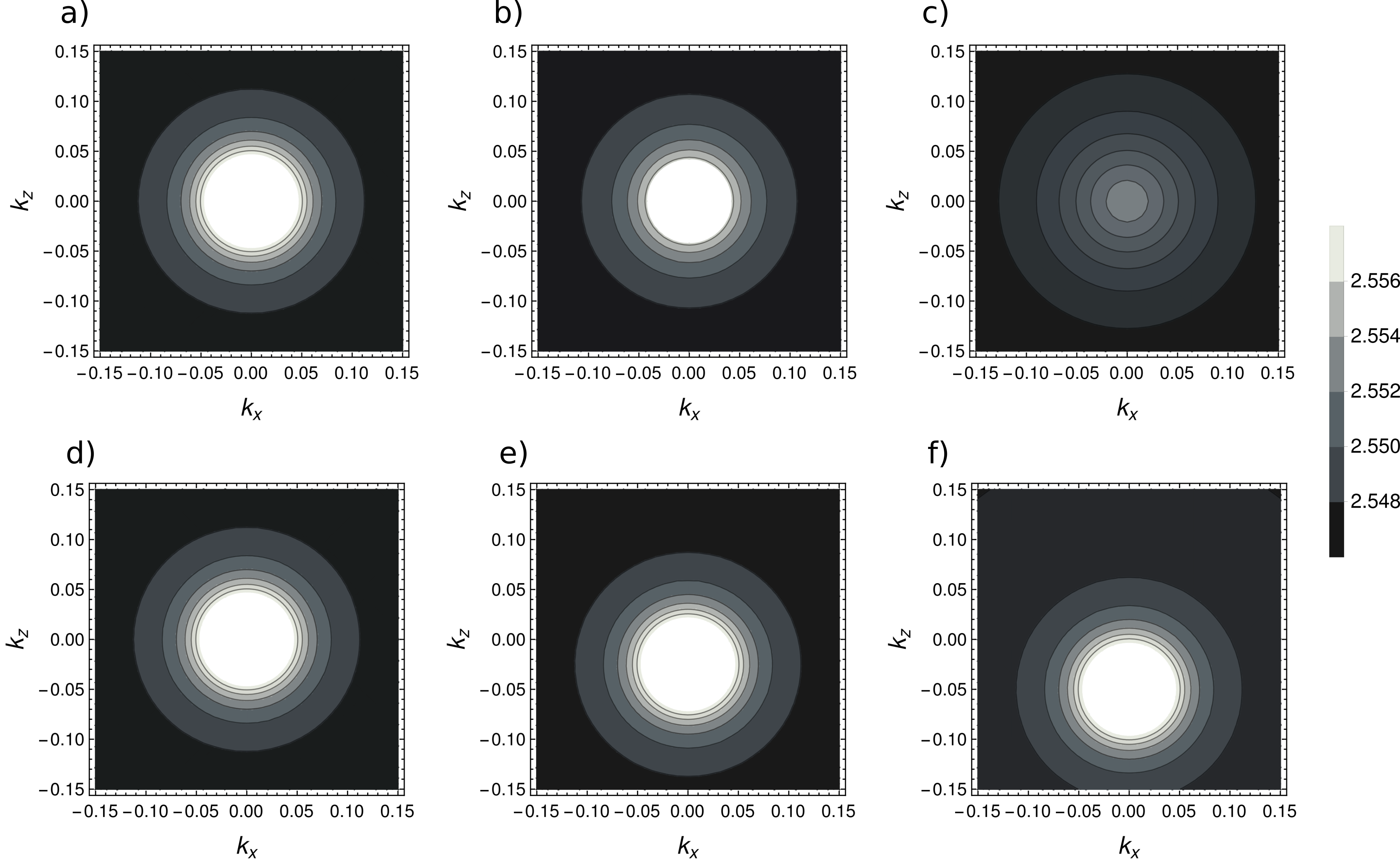}
  \caption{Contour plots of the densities of states in the $k_x - k_z$ plane (with $k_y=0$) at a constant energy, $E$, slightly off zero. Plots from left to right correspond to increasing amplitude of the applied light for $\phi=0$ [(a)-(c)] and for $\phi=\pi/2$ [(d)-(f)]. Plots in the top (bottom) row clearly show the dependence of the bandgap (shift) on $\gamma$. In the top row, as $\gamma$ increases, the high DOS near the centre of the $k_x-k_z$ plane (representing the degeneracy) decreases and gradually vanishes. In the bottom row, however, as $\gamma$ increases, the high DOS region shifts along $k_z$. Here we choose the following values for the different parameters: $\theta=\pi/2$, $\omega=30$, $E=0.0025$, $\eta=0.00005$, $A_y=2$ and $A_x$ varying in integer steps from $0-2$ (left to right).}
  \label{fig:DOS}
\end{figure}

Having discussed the tuning of the band structure with light, we next illustrate these changes using the $\bk$-resolved density of states (DOS) which can be directly measured in angle resolved photoemission (ARPES) experiments. The DOS at an energy $E$ given by $\rho(E,\bk)$, can be calculated using~\cite{bruus2004many}

\begin{equation}
\rho(E,\bk)=-\frac{1}{\pi} \Im(\Tr(G(E,\bk))),
\end{equation}

where the energy dependent Green's function $G(E,\bk)$ is given by

\begin{equation}
G(E,\bk)=(E+i\eta-H_{floq}(\bk))^{-1} ,
\end{equation}

where $\eta \to 0^+$ is a positive infinitesimal. Contour plots for $\rho(E,\bk)$ obtained are presented in Figure~\ref{fig:DOS}. For the unperturbed system, we observe that the DOS for $E$ close to the Fermi energy is large near $\bk=0$ due to the degeneracy, but decreases gradually as we move away from it. (Note that exactly at $E=0$, the DOS would be very large due to the contributions from the flat band.)
When light is applied, the high density point representing the three fold degeneracy shifts along $k_z$ for $\phi=\pi/6$ (mod $\pi/3$). While, for other $\phi$ values, the high DOS near the center vanishes as the degeneracy is no longer present and a band gap opens up. 
The dependence of $\Delta_g$ and $k_z^0$ on the intensity of applied light can also be clearly seen from our results.

\subsection{Anomalous Hall conductivity}

The signatures of light-induced changes to three fold fermions could be observed in several experimental probes. We propose measurement of the anomalous Hall effect as a direct way to verify the light-induced changes. As we discussed before, for $\phi=\pi/6$ (mod $\pi/3$), we can write $H_{floq}(\bk)=H_{3f}(\bk-\delta\bk)$ where $\delta\bk=(0,0,\gamma\sin 3\phi)$. A band gap does not open up at these values of $\phi$, and  the 3-BTP continues to behave like a source or sink of Berry curvature with monopole charge $\pm 2$. Moreover, since the flat band is topologically trivial with Chern number zero, for these particular values of $\phi$, we can calculate the change in the anomalous Hall conductivity, $\Delta\sigma_{xy}$, at half filling using the relation~\cite{chiral_chiral}

\begin{equation}
    \Delta\sigma_{xy}=n \frac{e^2}{h} \frac{\nu_z}{2\pi}.
\end{equation}

Here $n$ is the Chern number of the lower band and $\nu_z$ is the shift of the 3-BTP along the $k_z$ direction after applying light. Here, we have restored factors of $\hbar$, $v_f$ and $e$ for ease of estimation. Thus, for our case we get,

\begin{equation}
     \Delta\sigma_{xy}=2 \times \frac{e^2}{h} \times \frac{\gamma\sin 3\phi}{2\pi}=\frac{e^2}{h} \frac{\gamma\sin 3\phi}{\pi}
     =\pm \frac{e^2\gamma}{\pi h}.
\end{equation}
As the monopole charge of our three fold fermion is twice that of a Weyl fermion, the resulting anomalous Hall signature also comes out to be double of its Weyl counterpart~\cite{3f_pi2_PhysRevB.100.165302}. We will next discuss the values of different parameters required for experimental verification of our proposal and show that they lie well within current reach.

\subsection{Experimental considerations}
\label{section:exp_cons}
Now, we estimate values for the change in Hall voltage and band gap likely to be measured in experiments. 
If our sample with transverse dimensions $l_x\cross l_y$ and thickness $l_z=d$ is illuminated with a high frequency laser of power $P$ shining normally along $z$ direction, the penetration depth for light would be given by $\delta(\omega)=\frac{n(\omega) c \epsilon_0}{\Re \sigma_{xx}(\omega)}$, where $n$ is the refractive index of the material, $\sigma_{xx}$ is the longitudinal Hall conductivity, $c$ is the speed of light and $\epsilon_0$ is the permittivity of free space. When a current $I_x$ is applied along $x$-direction, we can thus measure a Hall voltage $V_y$ along the perpendicular direction, given by~\cite{chiral_chiral,Weyl_NL_PhysRevLett.117.087402}

\begin{equation}
    V_y=\frac{\sigma_{xy}\delta/d}{\sigma_{xx}^2 + (\sigma_{xy}\delta/d)^2}\frac{I_x}{d}.
\end{equation}

As $\sigma_{xx}\gg \sigma_{xy}$, we can ignore the $\sigma_{xy}$ term in the denominator. The change in the Hall conductivity due to light will then read

\begin{equation}
    \Delta V_y \approx \frac{\Delta\sigma_{xy}\delta}{\sigma_{xx}^2 d}\frac{I_x}{d}.
\end{equation}

For circularly polarised light, $\sin\theta=1$ and $A_x=A_y=A$. Using $E\sim A\omega$, we thus get

\begin{equation}
    \gamma = \frac{e^2v_f A^2}{2\hbar^2\omega} =\frac{e^2\hbar v_f E^2}{2 (\hbar\omega)^3} = \frac{e^2\hbar v_f }{2 (\hbar\omega)^3} \frac{2P(1-R)}{l_xl_y n(\omega)c\epsilon_0},
\end{equation} 

where $R$ is the reflectivity of the sample. This yields

\begin{equation}
    \Delta V_y= \left( \frac{e^2}{\pi h} \frac{e^2\hbar v_f }{2 (\hbar\omega)^3} \frac{2P(1-R)}{l_xl_y n(\omega)c\epsilon_0}\right) \left( \frac{n(\omega) c \epsilon_0}{\Re \sigma_{xx}(\omega)}\right) \frac{I_x}{\sigma_{xx}^2d^2}.
\end{equation}

We choose the typical sample parameters as $l_x=l_y=100$ $\mu$m, $d=100$ nm, $\sigma_{xx}\sim10^6$ $\Omega^{-1}$m$^{-1}$, $\Re\sigma_{xx}(\omega)\sim 10^5$ $\Omega^{-1}$m$^{-1}$ and $v_f\sim 5\cross 10^5$ m/s and consider shining a mid-infrared laser pulse with $\omega=30$ THz and $P=1$ W on our sample. 
With $R=0.8$ and $I_x=1$ A, we finally obtain $\Delta V_y=290$ nV, which can be easily detected in experiment. 

As the band gap is directly proportional to the intensity of applied light $I\sim P/l_xl_y$, using a laser with the same power $P=1$ W but smaller spot size would lead to larger effect on the band structure. With $l_x=l_y=1$ $\mu$m, we get $\Delta_g=2\hbar v_f\gamma \approx 25$ meV, which can again be readily measured in current ARPES experiments.

Heating in Floquet systems is a major concern for experimental realisation. However, there are some known ways of evading the problem~\cite{AnnRevCondMatt}. Opening a gap above the conduction and below the valence bands~\cite{McIver2020} is a possible way of preventing heating. Additionally, Floquet phases can be supported in prethermal regimes, where heating takes a parametrically long time~\cite{prethermal1Kuwahara_2016,prethermal2Weidinger2017,prethermal3Abanin2017}. Finally, driven systems can be actively cooled by coupling to a bath. Under suitable conditions, Floquet phases can be stabilized in such systems~\cite{heating1PhysRevB.91.235133,heating2PhysRevB.91.155422,heating3PhysRevX.5.041050,heating4PhysRevB.97.014311}.

\section{Lattice Model}

Next, we consider a lattice model having two triply degenerate points of opposite monopole charge in the first Brillouin zone to investigate the nature of the Fermi arcs and the effect of light on them. 

We start with a simplified tight binding model for triple-component fermions without spin-orbit coupling (or $\phi=\pi/2$)~\cite{tb_PhysRevB.100.235201}. The Hamiltonian reads 

\begin{equation}
H(\bk) = t \sin k_x S_x + t \sin k_y S_y + [t_z \cos k_z+m(\cos k_x + \cos k_y - 2)] S_z,
\label{eq:hamtb_pi2}
\end{equation}

where $\pmb S=(S_x,S_y,S_z)$ is a spin-1 representation of SU(2)

\begin{equation}
\pmb S = i
\begin{bmatrix}
0 & \hat{e_x} & -\hat{e_y} \\
-\hat{e_x} & 0 & \hat{e_z} \\
\hat{e_y} & -\hat{e_z} & 0
\end{bmatrix}.
\label{eq:S_def}
\end{equation}

In the Hamiltonian, the $\cos k_z$ term vanishes at $k_z=\pm \pi/2$. We work with the case where $t_z=m$ such that the mass term can vanish at $(k_x,k_y)=(0,0)$ where the coefficients of $S_x$ and $S_y$ also go to zero. With the above constraint, the system possess two three fold degeneracies at $\bk=(0,0,\pm \frac{\pi}{2})$ in the band structure.
We generalize this lattice model for all values of $\phi$ and write the generalized Hamiltonian, $H_{tb}$, as

\begin{equation}
H_{tb}(\bk) =
\begin{bmatrix}
0 & e^{i\phi} t\sin k_x & e^{-i\phi} t\sin k_y \\
e^{-i\phi} t\sin k_x & 0 & e^{i\phi}m(\cos k_z +\cos k_x + \cos k_y-2) \\
e^{i\phi} t\sin k_y & e^{-i\phi}m(\cos k_z +\cos k_x + \cos k_y -2) & 0
\end{bmatrix}.
\label{eq:hamtb}
\end{equation}

\begin{figure}[h!]
  \centering
  \includegraphics[scale=0.48]{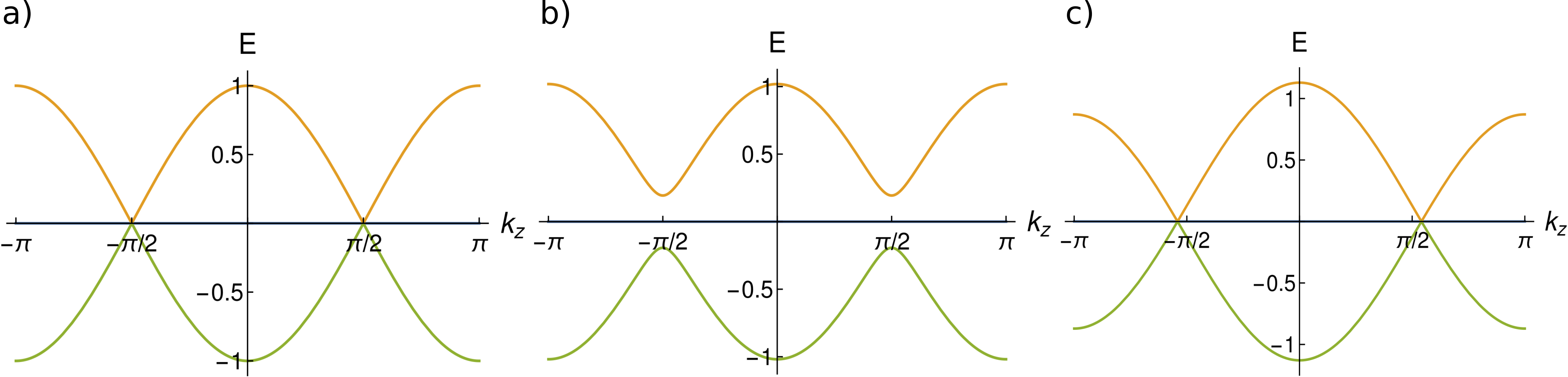}
  \caption{Band structure plots illustrating the effect of light on the lattice model. Eigenvalues of $H_{tb}$ (a) and $H_F$ (b,c) are as function of $k_z$ (at $k_x=k_y=0$). Plot (a) clearly shows the two three fold degenerate points at $\bk=(0,0,\pm \frac{\pi}{2})$ before applying light. 
  After applying light, a band gap opens up for $\phi=\pi/3$ (b), while the degeneracies shift along $k_z$ for $\phi=\pi/2$ (c). This matches our low energy model findings. Here we choose the following parameter values: $t=1$, $t_z=m=1$, $A_x=1$, $A_y=1$, and $\omega=3$..}
  \label{fig:lattice bands}
\end{figure}

The eigenvalues of $H_{tb}(\bk)$ are plotted in Figure~\ref{fig:lattice bands}(a) clearly showing the two 3-BTPs at $\bk=(0,0,\pm \frac{\pi}{2})$. We note that in the limit of $\bk \to (0,0,-\frac{\pi}{2})$, we recover exactly our low energy Hamiltonian, $H_{3f}$, as given by Equation~\ref{eq:ham0}. The band structure close to $k_z=\pm \frac{\pi}{2}$ also faithfully retains all the properties of the low energy Hamiltonian discussed above.

\subsection{Floquet analysis for the tight binding model}

Next, we examine the properties of the tight binding model upon illumination. When we shine light on our lattice system, we can again use Floquet theory to study the resulting effective band structures. 
From our analysis of the low energy model, we know that linearly polarised light cannot change the band structure while elliptically polarised light affects it the most. We thus choose $\theta=\pi/2$, giving the vector potential of light as $\pmb A=(A_x \cos \omega t,- A_y \sin \omega t,0)$. After some calculations, we find the lowest order light induced term to be

\begin{equation}
\frac{[V_{+1},V_{-1}]}{\omega} = i \zeta
\begin{bmatrix}
0 & -e^{-2i\phi}m t \sin k_x\cos k_y & e^{2i\phi}m t \sin k_y \cos k_x \\
e^{2i\phi}m t \sin k_x\cos k_y & 0 &-e^{-2i\phi}t^2 \cos k_x\cos k_y \\
-e^{-2i\phi}m t \sin k_y \cos k_x & e^{2i\phi}t^2 \cos k_x\cos k_y & 0
\end{bmatrix},
\label{eq:hamtb_floq}
\end{equation}

with $\zeta$ defined as

\begin{equation}
    \zeta=\frac{2 J_1(A_x)J_1(A_y)}{\omega},
\end{equation}

where $J_1$ is the Bessel function of the first kind.
The photon-dressed effective Hamiltonian reads $H_F=H_{tb} + \frac{[V_{+1},V_{-1}]}{\omega}$. Plots of the eigenvalues of $H_{F}$ are presented in Figures~\ref{fig:lattice bands}(b) and \ref{fig:lattice bands}(c). From the band structure, we observe that a band gap opens up and/or the two 3-BTPs shift along the $k_z$ axis depending on the value of $\phi$, similar to the case of the low energy model. The pattern exactly matches the one shown in Figure~\ref{variation_with_phi}, in agreement with our low energy effective model analysis. For $\phi \in (0,\frac{\pi}{3})$, the two 3-BTPs shift inward towards $k_z=0$, while for $\phi \in (\frac{\pi}{3},\frac{2\pi}{3})$, they shift outwards, and the pattern continues in a periodic fashion. 

\subsection{Surface States and Fermi Arcs}

\begin{figure}
    \centering
    \includegraphics[scale=0.5]{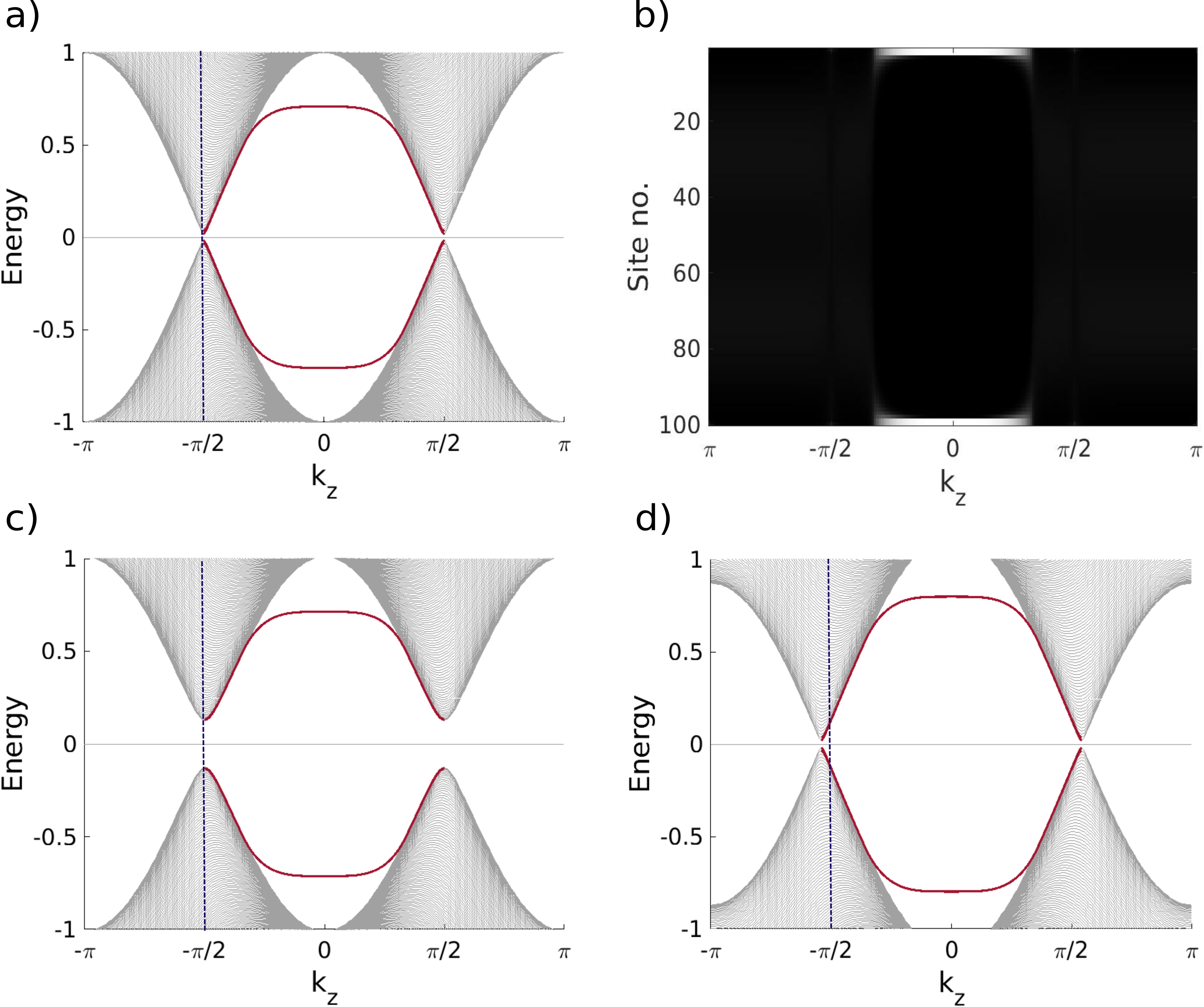}
    \caption{Numerically calculated band structure for the lattice model in a slab geometry (a) without light and (c)-(d) with light. We consider a slab with 200 sites along the $x$-direction having periodic boundary conditions along the $y$ and $z$ directions. All eigenvalues have been plotted along $k_z$ (at $k_y=0$). Surface states arising due to the non-zero Chern number of the bands have been highlighted in red. The sum of distributions of their corresponding wavefunctions is shown in (b). They are highly surface localised near $k_z=0$ but start penetrating into the bulk as the surface states mix with the bulk bands. Without application of light [panel (a)] the surface states connect the two three fold degeneracies. Under illumination with light, the surface states connect the shifted degeneracies for $\phi=\pi/2$ [panel (c)], while they do not traverse the bulk gap for $\phi=0$ [panel (d)]. Here we use the following parameter values: $t=1$, $t_z=m=1$, $A_x=1$, $A_y=1$, and $\omega=3$.}
    \label{fig:tb_plots}
\end{figure}

As our system has a non-trivial topology, we expect to observe topologically protected surface states representing the bulk-boundary correspondence. To investigate these surface states and the effect of light on them we consider our generalised tight binding model under open boundary conditions. We make our lattice finite along the $x$-direction, while keeping it periodic along the $y$ and $z$ directions. The numerical results for the band structure are shown in Figure~\ref{fig:tb_plots}(a). Apart from the bulk bands, we find that surface states connecting the two 3-BTPs appear in this slab geometry. As shown in Figure~\ref{fig:tb_plots}(b), their corresponding wavefunctions are highly localized near the surface for $k_z$ close to zero but start penetrating into the bulk as the surface states mix with the bulk bands. Both the upper and the lower bands contribute two such states, in agreement with the bulk Chern number for the two bands being $\pm 2$. 

After applying light, we observe that the surface states remain intact [see Figures~\ref{fig:tb_plots}(c) and \ref{fig:tb_plots}(d)], indicating that light does not change the Chern number of the bands. The structure of the surface states remains preserved for $\phi=\pi/6$ (mod $\pi/3$), however, they now connect the shifted 3-BTPs. 
For other values of $\phi$ where a gap opens up, the surface states do not traverse the bulk gap. The wavefunction for these surface states, however, does not show any significant change with respect to surface localization and decay upon illumination with light.

\subsection{Anomalous Hall conductivity}

\begin{figure}[h!]
  \centering
  \includegraphics[scale=0.8]{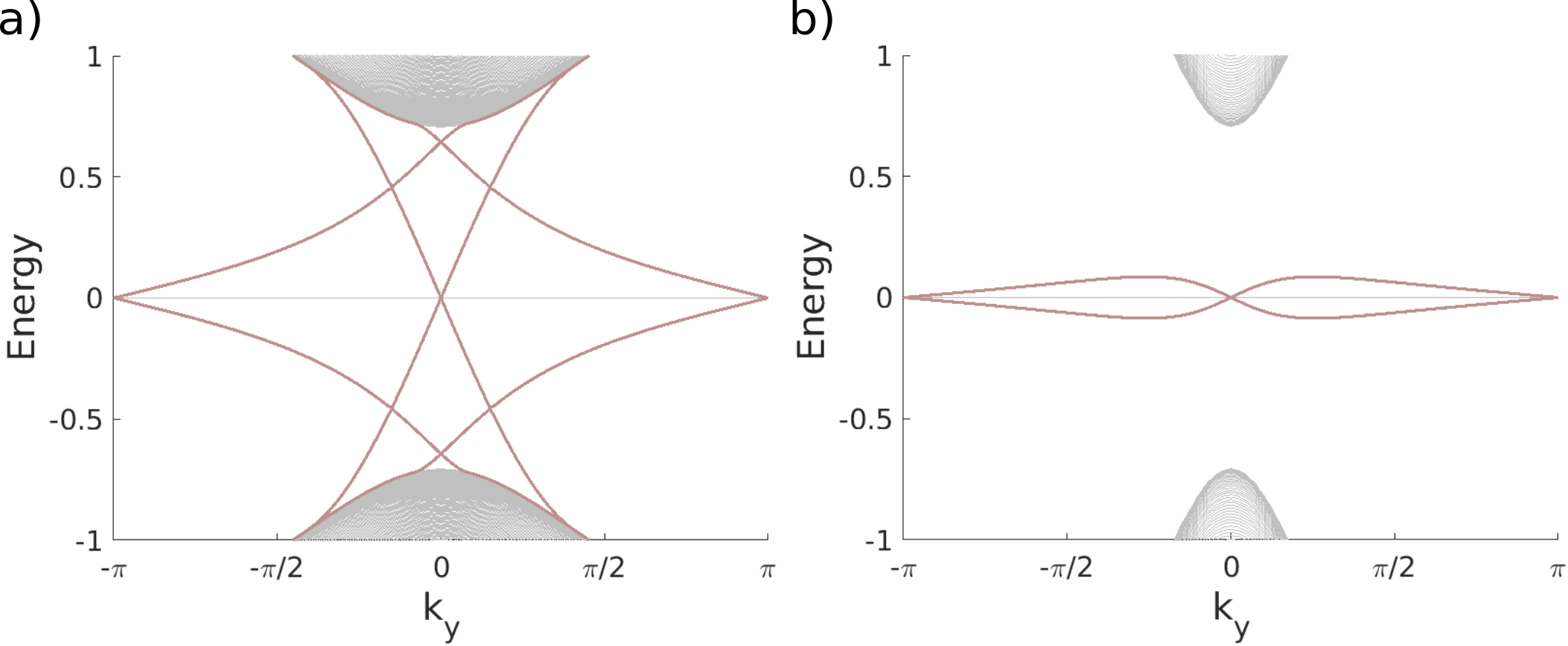}
  \caption{Illustration of the anomalous Hall insulator planes that stack to give rise to multifold fermion semimetal. (a) A plane at $k_z=\pi/4$ representing all planes in $k_z\in(-\pi/,\pi/2)$. (b) A plane at $k_z=3\pi/4$ representing all planes in $k_z\in(-\pi,-\pi/2) \cup (\pi/2,\pi)$. The band structure has been numerically calculated for the lattice model in a slab geometry similar to Figure \ref{fig:tb_plots}. Surface states arising due to the slab geometry are highlighted in red. In panel (a) we can see chiral surface states traversing the gap, whereas in panel (b) no such surface states are present. As before, we choose the following parameter values: $t=1$, $t_z=m=1$, $A_x=1$, $A_y=1$, and $\omega=3$.}
  \label{fig:2dcut}
\end{figure}

We now move on to calculate the change in the anomalous Hall signature for our lattice model for $\phi=\pi/2$ and compare it to our low energy effective model results.
Analogous to a Weyl semimetal~\cite{3dRev_RevModPhys.90.015001,bernevig2013topological}, our three fold semimetal can be thought of as being constructed by stacking planes of anomalous Hall insulators with Chern numbers $\pm 2$ along $k_z$~\cite{tb_PhysRevB.100.235201}. We show two such anomalous Hall insulator planes in Figure~\ref{fig:2dcut}. For $k_z\in(-\pi/2,\pi/2)$, each plane possesses chiral surface states traversing the gap which show a quantized Hall conductance. However, for planes beyond $\pi/2$, no such surface states are present. Thus, at half filling the semimetal would show a large anomalous Hall conductivity coming from contributions from each of the planes between the two 3-BTPs implying $\sigma_{xy}= 2\times \frac{e^2}{h}\times \frac{K_0}{2\pi}$ where $K_0$ is the separation between the two 3-BTPs. This gives $\sigma_{xy}=\frac{e^2}{h}$.

When light is applied, $K_0$ changes, leading to a change in the anomalous Hall signature.
To get an analytical form for the 3-BTP shift, we consider the mass term in $H_{tb}$ at $k_x=k_y=0$ which reads $\cos k_z+\zeta$. For small $\zeta$, the mass term vanishes for $k_z=\pm(\pi/2+\delta k_z)$. From the condition, $-\zeta=\cos(\pi/2+\delta k_z)$, where $\cos(\pi/2+\delta k_z)\approx -\delta k_z^2/2$, we thus obtain $\delta k_z=\zeta$. This tells us that the Hall signature changes by

\begin{equation}
    \Delta \sigma_{xy}=2\times \frac{e^2}{h}\times \frac{2\zeta}{2\pi}=\frac{2e^2\zeta}{\pi h}.
\end{equation}

The contribution from each node $\sim \frac{e^2\zeta}{\pi h}$ is similar to the low energy model results. Such a change in the anomalous Hall conductivity of the three fold fermion material upon illumination could be measured directly to test our predictions.

\section{Summary}

We have studied the effect of light on triple fold fermions, employing an effective low energy model and a simplified tight binding model, using the Floquet-Magnus expansion. We have shown that elliptically  polarized light can cause a band gap to open up and/or shift the three fold degeneracy depending on the intrinsic symmetries of the system. 
We show how the shift in the degeneracy changes the anomalous Hall conductivity causing an experimentally detectable change in the Hall voltage. We have also discussed the experimental feasibility of our proposals. We hope that our work further motivates exploration of light-matter interaction in multifold fermions.

\section*{Acknowledgments}

We thank Barry Bradlyn (University of Illinois at Urbana-Champaign) and Snehasish Nandy (University of Virginia) for useful discussions. R.J. thanks the Kishore Vaigyanik Protsahan Yojana (KVPY) for a fellowship. A.N. acknowledges support from the start-up grant (SG/MHRD-19-0001) of the Indian Institute of Science.

\appendix

\section{Validity of Floquet approximation}
\label{section:A_a}

In the Floquet expansion for the low energy model, the various terms in increasing powers of $(1/\omega)$ have the form: $H_{0}$, $V\frac{V}{\hbar\omega}$, $V(\frac{V}{\hbar\omega})^2$, and so on. Keeping terms only up to first order in $V/\hbar\omega$ is reasonable if $\omega \gg \frac{V}{\hbar} \approx \frac{e v_f A}{\hbar}$. 
Using $E\sim A\omega$, where $E$ is the root-mean-squared value of the electric field of light, we get the condition $\omega\gg\frac{e v_f E}{\hbar\omega}$ or $\omega \gg \sqrt{\frac{e v_f E}{\hbar}}$. Now, $E\sim\sqrt{\frac{2 I (1-R)}{n c \epsilon_0}}$, where, $I$ is the intensity of applied light. With the different parameters as given in Section \ref{section:exp_cons}, we get the condition on the possible applied frequencies as $\omega \gg 1$ THz.

\section{Analytical form of the energy bands}
\label{section:A_b}

The eigenvalues for $H_{floq}$ read
\begin{equation}
\begin{aligned}
& E_+ =\frac{2\hbar v_f}{\sqrt{3}}\sqrt{\lambda_1}\ \cos\left( \frac{\psi-\pi}{3} \right)
\quad
E_- = -\frac{2\hbar v_f}{\sqrt{3}} \sqrt{\lambda_1}\ \cos\left(  \frac{\psi}{3} \right)
\quad
E_0= \frac{2\hbar v_f}{\sqrt{3}} \sqrt{\lambda_1}\ \cos\left( \frac{\psi+\pi}{3} \right)\\
\text{with,} \quad
& \psi = 
     \begin{cases}
       \tan^{-1} \left(\frac{\sqrt{4\lambda_1^3-\lambda_2^2}}{\lambda_2}\right) &\quad \text{if}\ \lambda_2>0 \\
       \pi/2 &\quad \text{if}\ \lambda_2=0\\
       \tan^{-1} \left(\frac{\sqrt{4\lambda_1^3-\lambda_2^2}}{\lambda_2}\right) +\pi &\quad \text{if}\ \lambda_2<0\\
     \end{cases}\\
& \text{where,} \quad 
\lambda_1=\abs{k}^2+\gamma^2-2 k_z \gamma \sin 3\phi
\quad \text{and} \quad
\lambda_2=-54 k_x k_y k_z \cos 3\phi
\end{aligned}
\end{equation}

At special values of $\phi=\frac{\pi}{6}$ (mod $\pi/3$), we recover the eigenvalues $E_\pm=0, \pm \hbar v_f \sqrt{k_x^2+k_y^2+(k_z-\gamma/\hbar)^2}$ as discussed in the text.


\begin{thebibliography}{54}%
\makeatletter
\providecommand \@ifxundefined [1]{%
 \@ifx{#1\undefined}
}%
\providecommand \@ifnum [1]{%
 \ifnum #1\expandafter \@firstoftwo
 \else \expandafter \@secondoftwo
 \fi
}%
\providecommand \@ifx [1]{%
 \ifx #1\expandafter \@firstoftwo
 \else \expandafter \@secondoftwo
 \fi
}%
\providecommand \natexlab [1]{#1}%
\providecommand \enquote  [1]{``#1''}%
\providecommand \bibnamefont  [1]{#1}%
\providecommand \bibfnamefont [1]{#1}%
\providecommand \citenamefont [1]{#1}%
\providecommand \href@noop [0]{\@secondoftwo}%
\providecommand \href [0]{\begingroup \@sanitize@url \@href}%
\providecommand \@href[1]{\@@startlink{#1}\@@href}%
\providecommand \@@href[1]{\endgroup#1\@@endlink}%
\providecommand \@sanitize@url [0]{\catcode `\\12\catcode `\$12\catcode
  `\&12\catcode `\#12\catcode `\^12\catcode `\_12\catcode `\%12\relax}%
\providecommand \@@startlink[1]{}%
\providecommand \@@endlink[0]{}%
\providecommand \url  [0]{\begingroup\@sanitize@url \@url }%
\providecommand \@url [1]{\endgroup\@href {#1}{\urlprefix }}%
\providecommand \urlprefix  [0]{URL }%
\providecommand \Eprint [0]{\href }%
\providecommand \doibase [0]{http://dx.doi.org/}%
\providecommand \selectlanguage [0]{\@gobble}%
\providecommand \bibinfo  [0]{\@secondoftwo}%
\providecommand \bibfield  [0]{\@secondoftwo}%
\providecommand \translation [1]{[#1]}%
\providecommand \BibitemOpen [0]{}%
\providecommand \bibitemStop [0]{}%
\providecommand \bibitemNoStop [0]{.\EOS\space}%
\providecommand \EOS [0]{\spacefactor3000\relax}%
\providecommand \BibitemShut  [1]{\csname bibitem#1\endcsname}%
\let\auto@bib@innerbib\@empty
\bibitem [{\citenamefont {Klitzing}\ \emph {et~al.}(1980)\citenamefont
  {Klitzing}, \citenamefont {Dorda},\ and\ \citenamefont
  {Pepper}}]{QHE_Exp_PhysRevLett.45.494}%
  \BibitemOpen
  \bibfield  {author} {\bibinfo {author} {\bibfnamefont {K.~v.}\ \bibnamefont
  {Klitzing}}, \bibinfo {author} {\bibfnamefont {G.}~\bibnamefont {Dorda}}, \
  and\ \bibinfo {author} {\bibfnamefont {M.}~\bibnamefont {Pepper}},\ }\href
  {\doibase 10.1103/PhysRevLett.45.494} {\bibfield  {journal} {\bibinfo
  {journal} {Phys. Rev. Lett.}\ }\textbf {\bibinfo {volume} {45}},\ \bibinfo
  {pages} {494} (\bibinfo {year} {1980})}\BibitemShut {NoStop}%
\bibitem [{\citenamefont {Thouless}\ \emph {et~al.}(1982)\citenamefont
  {Thouless}, \citenamefont {Kohmoto}, \citenamefont {Nightingale},\ and\
  \citenamefont {den Nijs}}]{QHE_PhysRevLett.49.405}%
  \BibitemOpen
  \bibfield  {author} {\bibinfo {author} {\bibfnamefont {D.~J.}\ \bibnamefont
  {Thouless}}, \bibinfo {author} {\bibfnamefont {M.}~\bibnamefont {Kohmoto}},
  \bibinfo {author} {\bibfnamefont {M.~P.}\ \bibnamefont {Nightingale}}, \ and\
  \bibinfo {author} {\bibfnamefont {M.}~\bibnamefont {den Nijs}},\ }\href
  {\doibase 10.1103/PhysRevLett.49.405} {\bibfield  {journal} {\bibinfo
  {journal} {Phys. Rev. Lett.}\ }\textbf {\bibinfo {volume} {49}},\ \bibinfo
  {pages} {405} (\bibinfo {year} {1982})}\BibitemShut {NoStop}%
\bibitem [{\citenamefont {Hasan}\ and\ \citenamefont
  {Kane}(2010)}]{TI_RevModPhys.82.3045}%
  \BibitemOpen
  \bibfield  {author} {\bibinfo {author} {\bibfnamefont {M.~Z.}\ \bibnamefont
  {Hasan}}\ and\ \bibinfo {author} {\bibfnamefont {C.~L.}\ \bibnamefont
  {Kane}},\ }\href {\doibase 10.1103/RevModPhys.82.3045} {\bibfield  {journal}
  {\bibinfo  {journal} {Rev. Mod. Phys.}\ }\textbf {\bibinfo {volume} {82}},\
  \bibinfo {pages} {3045} (\bibinfo {year} {2010})}\BibitemShut {NoStop}%
\bibitem [{\citenamefont {Qi}\ and\ \citenamefont
  {Zhang}(2011)}]{TIS_RevModPhys.83.1057}%
  \BibitemOpen
  \bibfield  {author} {\bibinfo {author} {\bibfnamefont {X.-L.}\ \bibnamefont
  {Qi}}\ and\ \bibinfo {author} {\bibfnamefont {S.-C.}\ \bibnamefont {Zhang}},\
  }\href {\doibase 10.1103/RevModPhys.83.1057} {\bibfield  {journal} {\bibinfo
  {journal} {Rev. Mod. Phys.}\ }\textbf {\bibinfo {volume} {83}},\ \bibinfo
  {pages} {1057} (\bibinfo {year} {2011})}\BibitemShut {NoStop}%
\bibitem [{\citenamefont {Burkov}(2016)}]{Burkov2016}%
  \BibitemOpen
  \bibfield  {author} {\bibinfo {author} {\bibfnamefont {A.~A.}\ \bibnamefont
  {Burkov}},\ }\href {\doibase 10.1038/nmat4788} {\bibfield  {journal}
  {\bibinfo  {journal} {Nature Materials}\ }\textbf {\bibinfo {volume} {15}},\
  \bibinfo {pages} {1145} (\bibinfo {year} {2016})}\BibitemShut {NoStop}%
\bibitem [{\citenamefont {Armitage}\ \emph {et~al.}(2018)\citenamefont
  {Armitage}, \citenamefont {Mele},\ and\ \citenamefont
  {Vishwanath}}]{3dRev_RevModPhys.90.015001}%
  \BibitemOpen
  \bibfield  {author} {\bibinfo {author} {\bibfnamefont {N.~P.}\ \bibnamefont
  {Armitage}}, \bibinfo {author} {\bibfnamefont {E.~J.}\ \bibnamefont {Mele}},
  \ and\ \bibinfo {author} {\bibfnamefont {A.}~\bibnamefont {Vishwanath}},\
  }\href {\doibase 10.1103/RevModPhys.90.015001} {\bibfield  {journal}
  {\bibinfo  {journal} {Rev. Mod. Phys.}\ }\textbf {\bibinfo {volume} {90}},\
  \bibinfo {pages} {015001} (\bibinfo {year} {2018})}\BibitemShut {NoStop}%
\bibitem [{\citenamefont {Bradlyn}\ \emph {et~al.}(2016)\citenamefont
  {Bradlyn}, \citenamefont {Cano}, \citenamefont {Wang}, \citenamefont
  {Vergniory}, \citenamefont {Felser}, \citenamefont {Cava},\ and\
  \citenamefont {Bernevig}}]{Bradlynaaf5037}%
  \BibitemOpen
  \bibfield  {author} {\bibinfo {author} {\bibfnamefont {B.}~\bibnamefont
  {Bradlyn}}, \bibinfo {author} {\bibfnamefont {J.}~\bibnamefont {Cano}},
  \bibinfo {author} {\bibfnamefont {Z.}~\bibnamefont {Wang}}, \bibinfo {author}
  {\bibfnamefont {M.~G.}\ \bibnamefont {Vergniory}}, \bibinfo {author}
  {\bibfnamefont {C.}~\bibnamefont {Felser}}, \bibinfo {author} {\bibfnamefont
  {R.~J.}\ \bibnamefont {Cava}}, \ and\ \bibinfo {author} {\bibfnamefont
  {B.~A.}\ \bibnamefont {Bernevig}},\ }\href {\doibase 10.1126/science.aaf5037}
  {\ \textbf {\bibinfo {volume} {353}} (\bibinfo {year} {2016}),\
  10.1126/science.aaf5037}\BibitemShut {NoStop}%
\bibitem [{\citenamefont {Chang}\ \emph {et~al.}(2017)\citenamefont {Chang},
  \citenamefont {Xu}, \citenamefont {Wieder}, \citenamefont {Sanchez},
  \citenamefont {Huang}, \citenamefont {Belopolski}, \citenamefont {Chang},
  \citenamefont {Zhang}, \citenamefont {Bansil}, \citenamefont {Lin},\ and\
  \citenamefont {Hasan}}]{RhSi_PhysRevLett.119.206401}%
  \BibitemOpen
  \bibfield  {author} {\bibinfo {author} {\bibfnamefont {G.}~\bibnamefont
  {Chang}}, \bibinfo {author} {\bibfnamefont {S.-Y.}\ \bibnamefont {Xu}},
  \bibinfo {author} {\bibfnamefont {B.~J.}\ \bibnamefont {Wieder}}, \bibinfo
  {author} {\bibfnamefont {D.~S.}\ \bibnamefont {Sanchez}}, \bibinfo {author}
  {\bibfnamefont {S.-M.}\ \bibnamefont {Huang}}, \bibinfo {author}
  {\bibfnamefont {I.}~\bibnamefont {Belopolski}}, \bibinfo {author}
  {\bibfnamefont {T.-R.}\ \bibnamefont {Chang}}, \bibinfo {author}
  {\bibfnamefont {S.}~\bibnamefont {Zhang}}, \bibinfo {author} {\bibfnamefont
  {A.}~\bibnamefont {Bansil}}, \bibinfo {author} {\bibfnamefont
  {H.}~\bibnamefont {Lin}}, \ and\ \bibinfo {author} {\bibfnamefont {M.~Z.}\
  \bibnamefont {Hasan}},\ }\href {\doibase 10.1103/PhysRevLett.119.206401}
  {\bibfield  {journal} {\bibinfo  {journal} {Phys. Rev. Lett.}\ }\textbf
  {\bibinfo {volume} {119}},\ \bibinfo {pages} {206401} (\bibinfo {year}
  {2017})}\BibitemShut {NoStop}%
\bibitem [{\citenamefont {Takane}\ \emph {et~al.}(2019)\citenamefont {Takane},
  \citenamefont {Wang}, \citenamefont {Souma}, \citenamefont {Nakayama},
  \citenamefont {Nakamura}, \citenamefont {Oinuma}, \citenamefont {Nakata},
  \citenamefont {Iwasawa}, \citenamefont {Cacho}, \citenamefont {Kim},
  \citenamefont {Horiba}, \citenamefont {Kumigashira}, \citenamefont
  {Takahashi}, \citenamefont {Ando},\ and\ \citenamefont
  {Sato}}]{CoSi_PhysRevLett.122.076402}%
  \BibitemOpen
  \bibfield  {author} {\bibinfo {author} {\bibfnamefont {D.}~\bibnamefont
  {Takane}}, \bibinfo {author} {\bibfnamefont {Z.}~\bibnamefont {Wang}},
  \bibinfo {author} {\bibfnamefont {S.}~\bibnamefont {Souma}}, \bibinfo
  {author} {\bibfnamefont {K.}~\bibnamefont {Nakayama}}, \bibinfo {author}
  {\bibfnamefont {T.}~\bibnamefont {Nakamura}}, \bibinfo {author}
  {\bibfnamefont {H.}~\bibnamefont {Oinuma}}, \bibinfo {author} {\bibfnamefont
  {Y.}~\bibnamefont {Nakata}}, \bibinfo {author} {\bibfnamefont
  {H.}~\bibnamefont {Iwasawa}}, \bibinfo {author} {\bibfnamefont
  {C.}~\bibnamefont {Cacho}}, \bibinfo {author} {\bibfnamefont
  {T.}~\bibnamefont {Kim}}, \bibinfo {author} {\bibfnamefont {K.}~\bibnamefont
  {Horiba}}, \bibinfo {author} {\bibfnamefont {H.}~\bibnamefont {Kumigashira}},
  \bibinfo {author} {\bibfnamefont {T.}~\bibnamefont {Takahashi}}, \bibinfo
  {author} {\bibfnamefont {Y.}~\bibnamefont {Ando}}, \ and\ \bibinfo {author}
  {\bibfnamefont {T.}~\bibnamefont {Sato}},\ }\href {\doibase
  10.1103/PhysRevLett.122.076402} {\bibfield  {journal} {\bibinfo  {journal}
  {Phys. Rev. Lett.}\ }\textbf {\bibinfo {volume} {122}},\ \bibinfo {pages}
  {076402} (\bibinfo {year} {2019})}\BibitemShut {NoStop}%
\bibitem [{\citenamefont {Rao}\ \emph {et~al.}(2019)\citenamefont {Rao},
  \citenamefont {Li}, \citenamefont {Zhang}, \citenamefont {Tian},
  \citenamefont {Li}, \citenamefont {Fu}, \citenamefont {Tang}, \citenamefont
  {Wang}, \citenamefont {Li}, \citenamefont {Fan}, \citenamefont {Li},
  \citenamefont {Huang}, \citenamefont {Liu}, \citenamefont {Long},
  \citenamefont {Fang}, \citenamefont {Weng}, \citenamefont {Shi},
  \citenamefont {Lei}, \citenamefont {Sun}, \citenamefont {Qian},\ and\
  \citenamefont {Ding}}]{CoSi_Rao2019}%
  \BibitemOpen
  \bibfield  {author} {\bibinfo {author} {\bibfnamefont {Z.}~\bibnamefont
  {Rao}}, \bibinfo {author} {\bibfnamefont {H.}~\bibnamefont {Li}}, \bibinfo
  {author} {\bibfnamefont {T.}~\bibnamefont {Zhang}}, \bibinfo {author}
  {\bibfnamefont {S.}~\bibnamefont {Tian}}, \bibinfo {author} {\bibfnamefont
  {C.}~\bibnamefont {Li}}, \bibinfo {author} {\bibfnamefont {B.}~\bibnamefont
  {Fu}}, \bibinfo {author} {\bibfnamefont {C.}~\bibnamefont {Tang}}, \bibinfo
  {author} {\bibfnamefont {L.}~\bibnamefont {Wang}}, \bibinfo {author}
  {\bibfnamefont {Z.}~\bibnamefont {Li}}, \bibinfo {author} {\bibfnamefont
  {W.}~\bibnamefont {Fan}}, \bibinfo {author} {\bibfnamefont {J.}~\bibnamefont
  {Li}}, \bibinfo {author} {\bibfnamefont {Y.}~\bibnamefont {Huang}}, \bibinfo
  {author} {\bibfnamefont {Z.}~\bibnamefont {Liu}}, \bibinfo {author}
  {\bibfnamefont {Y.}~\bibnamefont {Long}}, \bibinfo {author} {\bibfnamefont
  {C.}~\bibnamefont {Fang}}, \bibinfo {author} {\bibfnamefont {H.}~\bibnamefont
  {Weng}}, \bibinfo {author} {\bibfnamefont {Y.}~\bibnamefont {Shi}}, \bibinfo
  {author} {\bibfnamefont {H.}~\bibnamefont {Lei}}, \bibinfo {author}
  {\bibfnamefont {Y.}~\bibnamefont {Sun}}, \bibinfo {author} {\bibfnamefont
  {T.}~\bibnamefont {Qian}}, \ and\ \bibinfo {author} {\bibfnamefont
  {H.}~\bibnamefont {Ding}},\ }\href {\doibase 10.1038/s41586-019-1031-8}
  {\bibfield  {journal} {\bibinfo  {journal} {Nature}\ }\textbf {\bibinfo
  {volume} {567}},\ \bibinfo {pages} {496} (\bibinfo {year}
  {2019})}\BibitemShut {NoStop}%
\bibitem [{\citenamefont {Schr{\"o}ter}\ \emph {et~al.}(2019)\citenamefont
  {Schr{\"o}ter}, \citenamefont {Pei}, \citenamefont {Vergniory}, \citenamefont
  {Sun}, \citenamefont {Manna}, \citenamefont {de~Juan}, \citenamefont
  {Krieger}, \citenamefont {S{\"u}ss}, \citenamefont {Schmidt}, \citenamefont
  {Dudin}, \citenamefont {Bradlyn}, \citenamefont {Kim}, \citenamefont
  {Schmitt}, \citenamefont {Cacho}, \citenamefont {Felser}, \citenamefont
  {Strocov},\ and\ \citenamefont {Chen}}]{AlPt_Schroter2019}%
  \BibitemOpen
  \bibfield  {author} {\bibinfo {author} {\bibfnamefont {N.~B.~M.}\
  \bibnamefont {Schr{\"o}ter}}, \bibinfo {author} {\bibfnamefont
  {D.}~\bibnamefont {Pei}}, \bibinfo {author} {\bibfnamefont {M.~G.}\
  \bibnamefont {Vergniory}}, \bibinfo {author} {\bibfnamefont {Y.}~\bibnamefont
  {Sun}}, \bibinfo {author} {\bibfnamefont {K.}~\bibnamefont {Manna}}, \bibinfo
  {author} {\bibfnamefont {F.}~\bibnamefont {de~Juan}}, \bibinfo {author}
  {\bibfnamefont {J.~A.}\ \bibnamefont {Krieger}}, \bibinfo {author}
  {\bibfnamefont {V.}~\bibnamefont {S{\"u}ss}}, \bibinfo {author}
  {\bibfnamefont {M.}~\bibnamefont {Schmidt}}, \bibinfo {author} {\bibfnamefont
  {P.}~\bibnamefont {Dudin}}, \bibinfo {author} {\bibfnamefont
  {B.}~\bibnamefont {Bradlyn}}, \bibinfo {author} {\bibfnamefont {T.~K.}\
  \bibnamefont {Kim}}, \bibinfo {author} {\bibfnamefont {T.}~\bibnamefont
  {Schmitt}}, \bibinfo {author} {\bibfnamefont {C.}~\bibnamefont {Cacho}},
  \bibinfo {author} {\bibfnamefont {C.}~\bibnamefont {Felser}}, \bibinfo
  {author} {\bibfnamefont {V.~N.}\ \bibnamefont {Strocov}}, \ and\ \bibinfo
  {author} {\bibfnamefont {Y.}~\bibnamefont {Chen}},\ }\href {\doibase
  10.1038/s41567-019-0511-y} {\bibfield  {journal} {\bibinfo  {journal} {Nature
  Physics}\ }\textbf {\bibinfo {volume} {15}},\ \bibinfo {pages} {759}
  (\bibinfo {year} {2019})}\BibitemShut {NoStop}%
\bibitem [{\citenamefont {Kumar}\ \emph {et~al.}(2020)\citenamefont {Kumar},
  \citenamefont {Yao}, \citenamefont {Nayak}, \citenamefont {Vergniory},
  \citenamefont {Bannies}, \citenamefont {Wang}, \citenamefont {Schröter},
  \citenamefont {Strocov}, \citenamefont {Müchler}, \citenamefont {Shi},
  \citenamefont {Rienks}, \citenamefont {Mañes}, \citenamefont {Shekhar},
  \citenamefont {Parkin}, \citenamefont {Fink}, \citenamefont {Fecher},
  \citenamefont {Sun}, \citenamefont {Bernevig},\ and\ \citenamefont
  {Felser}}]{PdSb2_doi:10.1002/adma.201906046}%
  \BibitemOpen
  \bibfield  {author} {\bibinfo {author} {\bibfnamefont {N.}~\bibnamefont
  {Kumar}}, \bibinfo {author} {\bibfnamefont {M.}~\bibnamefont {Yao}}, \bibinfo
  {author} {\bibfnamefont {J.}~\bibnamefont {Nayak}}, \bibinfo {author}
  {\bibfnamefont {M.~G.}\ \bibnamefont {Vergniory}}, \bibinfo {author}
  {\bibfnamefont {J.}~\bibnamefont {Bannies}}, \bibinfo {author} {\bibfnamefont
  {Z.}~\bibnamefont {Wang}}, \bibinfo {author} {\bibfnamefont {N.~B.~M.}\
  \bibnamefont {Schröter}}, \bibinfo {author} {\bibfnamefont {V.~N.}\
  \bibnamefont {Strocov}}, \bibinfo {author} {\bibfnamefont {L.}~\bibnamefont
  {Müchler}}, \bibinfo {author} {\bibfnamefont {W.}~\bibnamefont {Shi}},
  \bibinfo {author} {\bibfnamefont {E.~D.~L.}\ \bibnamefont {Rienks}}, \bibinfo
  {author} {\bibfnamefont {J.~L.}\ \bibnamefont {Mañes}}, \bibinfo {author}
  {\bibfnamefont {C.}~\bibnamefont {Shekhar}}, \bibinfo {author} {\bibfnamefont
  {S.~S.~P.}\ \bibnamefont {Parkin}}, \bibinfo {author} {\bibfnamefont
  {J.}~\bibnamefont {Fink}}, \bibinfo {author} {\bibfnamefont {G.~H.}\
  \bibnamefont {Fecher}}, \bibinfo {author} {\bibfnamefont {Y.}~\bibnamefont
  {Sun}}, \bibinfo {author} {\bibfnamefont {B.~A.}\ \bibnamefont {Bernevig}}, \
  and\ \bibinfo {author} {\bibfnamefont {C.}~\bibnamefont {Felser}},\ }\href
  {\doibase 10.1002/adma.201906046} {\bibfield  {journal} {\bibinfo  {journal}
  {Advanced Materials}\ }\textbf {\bibinfo {volume} {32}},\ \bibinfo {pages}
  {1906046} (\bibinfo {year} {2020})}\BibitemShut {NoStop}%
\bibitem [{\citenamefont {S\'anchez-Mart\'{\i}nez}\ \emph
  {et~al.}(2019)\citenamefont {S\'anchez-Mart\'{\i}nez}, \citenamefont
  {de~Juan},\ and\ \citenamefont {Grushin}}]{LinCon_PhysRevB.99.155145}%
  \BibitemOpen
  \bibfield  {author} {\bibinfo {author} {\bibfnamefont {M.-A.}\ \bibnamefont
  {S\'anchez-Mart\'{\i}nez}}, \bibinfo {author} {\bibfnamefont
  {F.}~\bibnamefont {de~Juan}}, \ and\ \bibinfo {author} {\bibfnamefont
  {A.~G.}\ \bibnamefont {Grushin}},\ }\href {\doibase
  10.1103/PhysRevB.99.155145} {\bibfield  {journal} {\bibinfo  {journal} {Phys.
  Rev. B}\ }\textbf {\bibinfo {volume} {99}},\ \bibinfo {pages} {155145}
  (\bibinfo {year} {2019})}\BibitemShut {NoStop}%
\bibitem [{\citenamefont {Flicker}\ \emph {et~al.}(2018)\citenamefont
  {Flicker}, \citenamefont {de~Juan}, \citenamefont {Bradlyn}, \citenamefont
  {Morimoto}, \citenamefont {Vergniory},\ and\ \citenamefont
  {Grushin}}]{CPGE_PhysRevB.98.155145}%
  \BibitemOpen
  \bibfield  {author} {\bibinfo {author} {\bibfnamefont {F.}~\bibnamefont
  {Flicker}}, \bibinfo {author} {\bibfnamefont {F.}~\bibnamefont {de~Juan}},
  \bibinfo {author} {\bibfnamefont {B.}~\bibnamefont {Bradlyn}}, \bibinfo
  {author} {\bibfnamefont {T.}~\bibnamefont {Morimoto}}, \bibinfo {author}
  {\bibfnamefont {M.~G.}\ \bibnamefont {Vergniory}}, \ and\ \bibinfo {author}
  {\bibfnamefont {A.~G.}\ \bibnamefont {Grushin}},\ }\href {\doibase
  10.1103/PhysRevB.98.155145} {\bibfield  {journal} {\bibinfo  {journal} {Phys.
  Rev. B}\ }\textbf {\bibinfo {volume} {98}},\ \bibinfo {pages} {155145}
  (\bibinfo {year} {2018})}\BibitemShut {NoStop}%
\bibitem [{\citenamefont {Rees}\ \emph {et~al.}(2019)\citenamefont {Rees},
  \citenamefont {Manna}, \citenamefont {Lu}, \citenamefont {Morimoto},
  \citenamefont {Borrmann}, \citenamefont {Felser}, \citenamefont {Moore},
  \citenamefont {Torchinsky},\ and\ \citenamefont
  {Orenstein}}]{rees2019observation}%
  \BibitemOpen
  \bibfield  {author} {\bibinfo {author} {\bibfnamefont {D.}~\bibnamefont
  {Rees}}, \bibinfo {author} {\bibfnamefont {K.}~\bibnamefont {Manna}},
  \bibinfo {author} {\bibfnamefont {B.}~\bibnamefont {Lu}}, \bibinfo {author}
  {\bibfnamefont {T.}~\bibnamefont {Morimoto}}, \bibinfo {author}
  {\bibfnamefont {H.}~\bibnamefont {Borrmann}}, \bibinfo {author}
  {\bibfnamefont {C.}~\bibnamefont {Felser}}, \bibinfo {author} {\bibfnamefont
  {J.~E.}\ \bibnamefont {Moore}}, \bibinfo {author} {\bibfnamefont {D.~H.}\
  \bibnamefont {Torchinsky}}, \ and\ \bibinfo {author} {\bibfnamefont
  {J.}~\bibnamefont {Orenstein}},\ }\href@noop {} {\enquote {\bibinfo {title}
  {Observation of topological photocurrents in the chiral weyl semimetal
  rhsi},}\ } (\bibinfo {year} {2019}),\ \Eprint
  {http://arxiv.org/abs/1902.03230} {arXiv:1902.03230 [cond-mat.mes-hall]}
  \BibitemShut {NoStop}%
\bibitem [{\citenamefont {Ni}\ \emph {et~al.}(2020{\natexlab{a}})\citenamefont
  {Ni}, \citenamefont {Wang}, \citenamefont {Zhang}, \citenamefont {Pozo},
  \citenamefont {Xu}, \citenamefont {Han}, \citenamefont {Manna}, \citenamefont
  {Paglione}, \citenamefont {Felser}, \citenamefont {Grushin}, \citenamefont
  {de~Juan}, \citenamefont {Mele},\ and\ \citenamefont {Wu}}]{ni2020giant}%
  \BibitemOpen
  \bibfield  {author} {\bibinfo {author} {\bibfnamefont {Z.}~\bibnamefont
  {Ni}}, \bibinfo {author} {\bibfnamefont {K.}~\bibnamefont {Wang}}, \bibinfo
  {author} {\bibfnamefont {Y.}~\bibnamefont {Zhang}}, \bibinfo {author}
  {\bibfnamefont {O.}~\bibnamefont {Pozo}}, \bibinfo {author} {\bibfnamefont
  {B.}~\bibnamefont {Xu}}, \bibinfo {author} {\bibfnamefont {X.}~\bibnamefont
  {Han}}, \bibinfo {author} {\bibfnamefont {K.}~\bibnamefont {Manna}}, \bibinfo
  {author} {\bibfnamefont {J.}~\bibnamefont {Paglione}}, \bibinfo {author}
  {\bibfnamefont {C.}~\bibnamefont {Felser}}, \bibinfo {author} {\bibfnamefont
  {A.~G.}\ \bibnamefont {Grushin}}, \bibinfo {author} {\bibfnamefont
  {F.}~\bibnamefont {de~Juan}}, \bibinfo {author} {\bibfnamefont {E.~J.}\
  \bibnamefont {Mele}}, \ and\ \bibinfo {author} {\bibfnamefont
  {L.}~\bibnamefont {Wu}},\ }\href@noop {} {\enquote {\bibinfo {title} {Giant
  topological longitudinal circular photo-galvanic effect in the chiral
  multifold semimetal cosi},}\ } (\bibinfo {year} {2020}{\natexlab{a}}),\
  \Eprint {http://arxiv.org/abs/2006.09612} {arXiv:2006.09612
  [cond-mat.mtrl-sci]} \BibitemShut {NoStop}%
\bibitem [{\citenamefont {Ni}\ \emph {et~al.}(2020{\natexlab{b}})\citenamefont
  {Ni}, \citenamefont {Xu}, \citenamefont {Sanchez-Martinez}, \citenamefont
  {Zhang}, \citenamefont {Manna}, \citenamefont {Bernhard}, \citenamefont
  {Venderbos}, \citenamefont {de~Juan}, \citenamefont {Felser}, \citenamefont
  {Grushin},\ and\ \citenamefont {Wu}}]{ni2020linear}%
  \BibitemOpen
  \bibfield  {author} {\bibinfo {author} {\bibfnamefont {Z.}~\bibnamefont
  {Ni}}, \bibinfo {author} {\bibfnamefont {B.}~\bibnamefont {Xu}}, \bibinfo
  {author} {\bibfnamefont {M.~A.}\ \bibnamefont {Sanchez-Martinez}}, \bibinfo
  {author} {\bibfnamefont {Y.}~\bibnamefont {Zhang}}, \bibinfo {author}
  {\bibfnamefont {K.}~\bibnamefont {Manna}}, \bibinfo {author} {\bibfnamefont
  {C.}~\bibnamefont {Bernhard}}, \bibinfo {author} {\bibfnamefont {J.~W.~F.}\
  \bibnamefont {Venderbos}}, \bibinfo {author} {\bibfnamefont {F.}~\bibnamefont
  {de~Juan}}, \bibinfo {author} {\bibfnamefont {C.}~\bibnamefont {Felser}},
  \bibinfo {author} {\bibfnamefont {A.~G.}\ \bibnamefont {Grushin}}, \ and\
  \bibinfo {author} {\bibfnamefont {L.}~\bibnamefont {Wu}},\ }\href@noop {}
  {\enquote {\bibinfo {title} {Linear and nonlinear optical responses in the
  chiral multifold semimetal rhsi},}\ } (\bibinfo {year}
  {2020}{\natexlab{b}}),\ \Eprint {http://arxiv.org/abs/2005.13473}
  {arXiv:2005.13473 [cond-mat.mtrl-sci]} \BibitemShut {NoStop}%
\bibitem [{\citenamefont {Cayssol}\ \emph {et~al.}(2013)\citenamefont
  {Cayssol}, \citenamefont {D{\'o}ra}, \citenamefont {Simon},\ and\
  \citenamefont {Moessner}}]{review_moessner}%
  \BibitemOpen
  \bibfield  {author} {\bibinfo {author} {\bibfnamefont {J.}~\bibnamefont
  {Cayssol}}, \bibinfo {author} {\bibfnamefont {B.}~\bibnamefont {D{\'o}ra}},
  \bibinfo {author} {\bibfnamefont {F.}~\bibnamefont {Simon}}, \ and\ \bibinfo
  {author} {\bibfnamefont {R.}~\bibnamefont {Moessner}},\ }\href {\doibase
  10.1002/pssr.201206451} {\bibfield  {journal} {\bibinfo  {journal} {Physica
  Status Solidi - Rapid Research Letters}\ }\textbf {\bibinfo {volume} {7}},\
  \bibinfo {pages} {101} (\bibinfo {year} {2013})}\BibitemShut {NoStop}%
\bibitem [{\citenamefont {Giovannini}\ and\ \citenamefont
  {Hübener}(2019)}]{RevGiovannini_2019}%
  \BibitemOpen
  \bibfield  {author} {\bibinfo {author} {\bibfnamefont {U.~D.}\ \bibnamefont
  {Giovannini}}\ and\ \bibinfo {author} {\bibfnamefont {H.}~\bibnamefont
  {Hübener}},\ }\href {\doibase 10.1088/2515-7639/ab387b} {\bibfield
  {journal} {\bibinfo  {journal} {Journal of Physics: Materials}\ }\textbf
  {\bibinfo {volume} {3}},\ \bibinfo {pages} {012001} (\bibinfo {year}
  {2019})}\BibitemShut {NoStop}%
\bibitem [{\citenamefont {Oka}\ and\ \citenamefont
  {Kitamura}(2019)}]{reviewoka}%
  \BibitemOpen
  \bibfield  {author} {\bibinfo {author} {\bibfnamefont {T.}~\bibnamefont
  {Oka}}\ and\ \bibinfo {author} {\bibfnamefont {S.}~\bibnamefont {Kitamura}},\
  }\href {\doibase 10.1146/annurev-conmatphys-031218-013423} {\bibfield
  {journal} {\bibinfo  {journal} {Annual Review of Condensed Matter Physics}\
  }\textbf {\bibinfo {volume} {10}},\ \bibinfo {pages} {387} (\bibinfo {year}
  {2019})},\ \Eprint
  {http://arxiv.org/abs/https://doi.org/10.1146/annurev-conmatphys-031218-013423}
  {https://doi.org/10.1146/annurev-conmatphys-031218-013423} \BibitemShut
  {NoStop}%
\bibitem [{\citenamefont {Oka}\ and\ \citenamefont {Aoki}(2009)}]{oka2009}%
  \BibitemOpen
  \bibfield  {author} {\bibinfo {author} {\bibfnamefont {T.}~\bibnamefont
  {Oka}}\ and\ \bibinfo {author} {\bibfnamefont {H.}~\bibnamefont {Aoki}},\
  }\href {\doibase 10.1103/PhysRevB.79.081406} {\bibfield  {journal} {\bibinfo
  {journal} {Phys. Rev. B}\ }\textbf {\bibinfo {volume} {79}},\ \bibinfo
  {pages} {081406} (\bibinfo {year} {2009})}\BibitemShut {NoStop}%
\bibitem [{\citenamefont {Lindner}\ \emph {et~al.}(2011)\citenamefont
  {Lindner}, \citenamefont {Refael},\ and\ \citenamefont
  {Galitski}}]{Lindner2011}%
  \BibitemOpen
  \bibfield  {author} {\bibinfo {author} {\bibfnamefont {N.~H.}\ \bibnamefont
  {Lindner}}, \bibinfo {author} {\bibfnamefont {G.}~\bibnamefont {Refael}}, \
  and\ \bibinfo {author} {\bibfnamefont {V.}~\bibnamefont {Galitski}},\ }\href
  {\doibase 10.1038/nphys1926} {\bibfield  {journal} {\bibinfo  {journal}
  {Nature Physics}\ }\textbf {\bibinfo {volume} {7}},\ \bibinfo {pages} {490}
  (\bibinfo {year} {2011})}\BibitemShut {NoStop}%
\bibitem [{\citenamefont {Kitagawa}\ \emph {et~al.}(2011)\citenamefont
  {Kitagawa}, \citenamefont {Oka}, \citenamefont {Brataas}, \citenamefont
  {Fu},\ and\ \citenamefont {Demler}}]{Oka2011}%
  \BibitemOpen
  \bibfield  {author} {\bibinfo {author} {\bibfnamefont {T.}~\bibnamefont
  {Kitagawa}}, \bibinfo {author} {\bibfnamefont {T.}~\bibnamefont {Oka}},
  \bibinfo {author} {\bibfnamefont {A.}~\bibnamefont {Brataas}}, \bibinfo
  {author} {\bibfnamefont {L.}~\bibnamefont {Fu}}, \ and\ \bibinfo {author}
  {\bibfnamefont {E.}~\bibnamefont {Demler}},\ }\href {\doibase
  10.1103/PhysRevB.84.235108} {\bibfield  {journal} {\bibinfo  {journal} {Phys.
  Rev. B}\ }\textbf {\bibinfo {volume} {84}},\ \bibinfo {pages} {235108}
  (\bibinfo {year} {2011})}\BibitemShut {NoStop}%
\bibitem [{\citenamefont {Chan}\ \emph
  {et~al.}(2016{\natexlab{a}})\citenamefont {Chan}, \citenamefont {Lee},
  \citenamefont {Burch}, \citenamefont {Han},\ and\ \citenamefont
  {Ran}}]{chiral_chiral}%
  \BibitemOpen
  \bibfield  {author} {\bibinfo {author} {\bibfnamefont {C.-K.}\ \bibnamefont
  {Chan}}, \bibinfo {author} {\bibfnamefont {P.~A.}\ \bibnamefont {Lee}},
  \bibinfo {author} {\bibfnamefont {K.~S.}\ \bibnamefont {Burch}}, \bibinfo
  {author} {\bibfnamefont {J.~H.}\ \bibnamefont {Han}}, \ and\ \bibinfo
  {author} {\bibfnamefont {Y.}~\bibnamefont {Ran}},\ }\href {\doibase
  10.1103/PhysRevLett.116.026805} {\bibfield  {journal} {\bibinfo  {journal}
  {Phys. Rev. Lett.}\ }\textbf {\bibinfo {volume} {116}},\ \bibinfo {pages}
  {026805} (\bibinfo {year} {2016}{\natexlab{a}})}\BibitemShut {NoStop}%
\bibitem [{\citenamefont {Ezawa}(2013)}]{TI_TI_PhysRevLett.110.026603}%
  \BibitemOpen
  \bibfield  {author} {\bibinfo {author} {\bibfnamefont {M.}~\bibnamefont
  {Ezawa}},\ }\href {\doibase 10.1103/PhysRevLett.110.026603} {\bibfield
  {journal} {\bibinfo  {journal} {Phys. Rev. Lett.}\ }\textbf {\bibinfo
  {volume} {110}},\ \bibinfo {pages} {026603} (\bibinfo {year}
  {2013})}\BibitemShut {NoStop}%
\bibitem [{\citenamefont {Narayan}(2015)}]{AN_PhysRevB.91.205445}%
  \BibitemOpen
  \bibfield  {author} {\bibinfo {author} {\bibfnamefont {A.}~\bibnamefont
  {Narayan}},\ }\href {\doibase 10.1103/PhysRevB.91.205445} {\bibfield
  {journal} {\bibinfo  {journal} {Phys. Rev. B}\ }\textbf {\bibinfo {volume}
  {91}},\ \bibinfo {pages} {205445} (\bibinfo {year} {2015})}\BibitemShut
  {NoStop}%
\bibitem [{\citenamefont {Wang}\ \emph {et~al.}(2013)\citenamefont {Wang},
  \citenamefont {Steinberg}, \citenamefont {Jarillo-Herrero},\ and\
  \citenamefont {Gedik}}]{Wang453}%
  \BibitemOpen
  \bibfield  {author} {\bibinfo {author} {\bibfnamefont {Y.~H.}\ \bibnamefont
  {Wang}}, \bibinfo {author} {\bibfnamefont {H.}~\bibnamefont {Steinberg}},
  \bibinfo {author} {\bibfnamefont {P.}~\bibnamefont {Jarillo-Herrero}}, \ and\
  \bibinfo {author} {\bibfnamefont {N.}~\bibnamefont {Gedik}},\ }\href
  {\doibase 10.1126/science.1239834} {\bibfield  {journal} {\bibinfo  {journal}
  {Science}\ }\textbf {\bibinfo {volume} {342}},\ \bibinfo {pages} {453}
  (\bibinfo {year} {2013})},\ \Eprint
  {http://arxiv.org/abs/https://science.sciencemag.org/content/342/6157/453.full.pdf}
  {https://science.sciencemag.org/content/342/6157/453.full.pdf} \BibitemShut
  {NoStop}%
\bibitem [{\citenamefont {McIver}\ \emph {et~al.}(2020)\citenamefont {McIver},
  \citenamefont {Schulte}, \citenamefont {Stein}, \citenamefont {Matsuyama},
  \citenamefont {Jotzu}, \citenamefont {Meier},\ and\ \citenamefont
  {Cavalleri}}]{McIver2020}%
  \BibitemOpen
  \bibfield  {author} {\bibinfo {author} {\bibfnamefont {J.~W.}\ \bibnamefont
  {McIver}}, \bibinfo {author} {\bibfnamefont {B.}~\bibnamefont {Schulte}},
  \bibinfo {author} {\bibfnamefont {F.-U.}\ \bibnamefont {Stein}}, \bibinfo
  {author} {\bibfnamefont {T.}~\bibnamefont {Matsuyama}}, \bibinfo {author}
  {\bibfnamefont {G.}~\bibnamefont {Jotzu}}, \bibinfo {author} {\bibfnamefont
  {G.}~\bibnamefont {Meier}}, \ and\ \bibinfo {author} {\bibfnamefont
  {A.}~\bibnamefont {Cavalleri}},\ }\href {\doibase 10.1038/s41567-019-0698-y}
  {\bibfield  {journal} {\bibinfo  {journal} {Nature Physics}\ }\textbf
  {\bibinfo {volume} {16}},\ \bibinfo {pages} {38} (\bibinfo {year}
  {2020})}\BibitemShut {NoStop}%
\bibitem [{\citenamefont {L{\'{o}}pez}\ and\ \citenamefont
  {Molina}(2020)}]{L_pez_2020}%
  \BibitemOpen
  \bibfield  {author} {\bibinfo {author} {\bibfnamefont {A.}~\bibnamefont
  {L{\'{o}}pez}}\ and\ \bibinfo {author} {\bibfnamefont {R.~A.}\ \bibnamefont
  {Molina}},\ }\href {\doibase 10.1088/1361-648x/ab6cc0} {\bibfield  {journal}
  {\bibinfo  {journal} {Journal of Physics: Condensed Matter}\ }\textbf
  {\bibinfo {volume} {32}},\ \bibinfo {pages} {205701} (\bibinfo {year}
  {2020})}\BibitemShut {NoStop}%
\bibitem [{\citenamefont {H{\"u}bener}\ \emph {et~al.}(2017)\citenamefont
  {H{\"u}bener}, \citenamefont {Sentef}, \citenamefont {De~Giovannini},
  \citenamefont {Kemper},\ and\ \citenamefont
  {Rubio}}]{Weyl_Dirac_Hubener2017}%
  \BibitemOpen
  \bibfield  {author} {\bibinfo {author} {\bibfnamefont {H.}~\bibnamefont
  {H{\"u}bener}}, \bibinfo {author} {\bibfnamefont {M.~A.}\ \bibnamefont
  {Sentef}}, \bibinfo {author} {\bibfnamefont {U.}~\bibnamefont
  {De~Giovannini}}, \bibinfo {author} {\bibfnamefont {A.~F.}\ \bibnamefont
  {Kemper}}, \ and\ \bibinfo {author} {\bibfnamefont {A.}~\bibnamefont
  {Rubio}},\ }\href {\doibase 10.1038/ncomms13940} {\bibfield  {journal}
  {\bibinfo  {journal} {Nature Communications}\ }\textbf {\bibinfo {volume}
  {8}},\ \bibinfo {pages} {13940} (\bibinfo {year} {2017})}\BibitemShut
  {NoStop}%
\bibitem [{\citenamefont {Wang}\ \emph {et~al.}(2014)\citenamefont {Wang},
  \citenamefont {Wang}, \citenamefont {Shen}, \citenamefont {Sheng},\ and\
  \citenamefont {Xing}}]{Weyl_TI_Wang_2014}%
  \BibitemOpen
  \bibfield  {author} {\bibinfo {author} {\bibfnamefont {R.}~\bibnamefont
  {Wang}}, \bibinfo {author} {\bibfnamefont {B.}~\bibnamefont {Wang}}, \bibinfo
  {author} {\bibfnamefont {R.}~\bibnamefont {Shen}}, \bibinfo {author}
  {\bibfnamefont {L.}~\bibnamefont {Sheng}}, \ and\ \bibinfo {author}
  {\bibfnamefont {D.~Y.}\ \bibnamefont {Xing}},\ }\href {\doibase
  10.1209/0295-5075/105/17004} {\bibfield  {journal} {\bibinfo  {journal}
  {{EPL} (Europhysics Letters)}\ }\textbf {\bibinfo {volume} {105}},\ \bibinfo
  {pages} {17004} (\bibinfo {year} {2014})}\BibitemShut {NoStop}%
\bibitem [{\citenamefont {Narayan}(2016)}]{Weyl_NL_PhysRevB.94.041409}%
  \BibitemOpen
  \bibfield  {author} {\bibinfo {author} {\bibfnamefont {A.}~\bibnamefont
  {Narayan}},\ }\href {\doibase 10.1103/PhysRevB.94.041409} {\bibfield
  {journal} {\bibinfo  {journal} {Phys. Rev. B}\ }\textbf {\bibinfo {volume}
  {94}},\ \bibinfo {pages} {041409} (\bibinfo {year} {2016})}\BibitemShut
  {NoStop}%
\bibitem [{\citenamefont {Yan}\ and\ \citenamefont
  {Wang}(2016)}]{Weyl_NL_PhysRevLett.117.087402}%
  \BibitemOpen
  \bibfield  {author} {\bibinfo {author} {\bibfnamefont {Z.}~\bibnamefont
  {Yan}}\ and\ \bibinfo {author} {\bibfnamefont {Z.}~\bibnamefont {Wang}},\
  }\href {\doibase 10.1103/PhysRevLett.117.087402} {\bibfield  {journal}
  {\bibinfo  {journal} {Phys. Rev. Lett.}\ }\textbf {\bibinfo {volume} {117}},\
  \bibinfo {pages} {087402} (\bibinfo {year} {2016})}\BibitemShut {NoStop}%
\bibitem [{\citenamefont {Chan}\ \emph
  {et~al.}(2016{\natexlab{b}})\citenamefont {Chan}, \citenamefont {Oh},
  \citenamefont {Han},\ and\ \citenamefont
  {Lee}}]{Weyl_NLD_PhysRevB.94.121106}%
  \BibitemOpen
  \bibfield  {author} {\bibinfo {author} {\bibfnamefont {C.-K.}\ \bibnamefont
  {Chan}}, \bibinfo {author} {\bibfnamefont {Y.-T.}\ \bibnamefont {Oh}},
  \bibinfo {author} {\bibfnamefont {J.~H.}\ \bibnamefont {Han}}, \ and\
  \bibinfo {author} {\bibfnamefont {P.~A.}\ \bibnamefont {Lee}},\ }\href
  {\doibase 10.1103/PhysRevB.94.121106} {\bibfield  {journal} {\bibinfo
  {journal} {Phys. Rev. B}\ }\textbf {\bibinfo {volume} {94}},\ \bibinfo
  {pages} {121106} (\bibinfo {year} {2016}{\natexlab{b}})}\BibitemShut
  {NoStop}%
\bibitem [{\citenamefont {Grushin}\ \emph {et~al.}(2014)\citenamefont
  {Grushin}, \citenamefont {G\'omez-Le\'on},\ and\ \citenamefont
  {Neupert}}]{fracCI_PhysRevLett.112.156801}%
  \BibitemOpen
  \bibfield  {author} {\bibinfo {author} {\bibfnamefont {A.~G.}\ \bibnamefont
  {Grushin}}, \bibinfo {author} {\bibfnamefont {A.}~\bibnamefont
  {G\'omez-Le\'on}}, \ and\ \bibinfo {author} {\bibfnamefont {T.}~\bibnamefont
  {Neupert}},\ }\href {\doibase 10.1103/PhysRevLett.112.156801} {\bibfield
  {journal} {\bibinfo  {journal} {Phys. Rev. Lett.}\ }\textbf {\bibinfo
  {volume} {112}},\ \bibinfo {pages} {156801} (\bibinfo {year}
  {2014})}\BibitemShut {NoStop}%
\bibitem [{\citenamefont {Takasan}\ \emph {et~al.}(2017)\citenamefont
  {Takasan}, \citenamefont {Daido}, \citenamefont {Kawakami},\ and\
  \citenamefont {Yanase}}]{supercond_PhysRevB.95.134508}%
  \BibitemOpen
  \bibfield  {author} {\bibinfo {author} {\bibfnamefont {K.}~\bibnamefont
  {Takasan}}, \bibinfo {author} {\bibfnamefont {A.}~\bibnamefont {Daido}},
  \bibinfo {author} {\bibfnamefont {N.}~\bibnamefont {Kawakami}}, \ and\
  \bibinfo {author} {\bibfnamefont {Y.}~\bibnamefont {Yanase}},\ }\href
  {\doibase 10.1103/PhysRevB.95.134508} {\bibfield  {journal} {\bibinfo
  {journal} {Phys. Rev. B}\ }\textbf {\bibinfo {volume} {95}},\ \bibinfo
  {pages} {134508} (\bibinfo {year} {2017})}\BibitemShut {NoStop}%
\bibitem [{\citenamefont {Topp}\ \emph {et~al.}(2019)\citenamefont {Topp},
  \citenamefont {Jotzu}, \citenamefont {McIver}, \citenamefont {Xian},
  \citenamefont {Rubio},\ and\ \citenamefont
  {Sentef}}]{blgraphene_PhysRevResearch.1.023031}%
  \BibitemOpen
  \bibfield  {author} {\bibinfo {author} {\bibfnamefont {G.~E.}\ \bibnamefont
  {Topp}}, \bibinfo {author} {\bibfnamefont {G.}~\bibnamefont {Jotzu}},
  \bibinfo {author} {\bibfnamefont {J.~W.}\ \bibnamefont {McIver}}, \bibinfo
  {author} {\bibfnamefont {L.}~\bibnamefont {Xian}}, \bibinfo {author}
  {\bibfnamefont {A.}~\bibnamefont {Rubio}}, \ and\ \bibinfo {author}
  {\bibfnamefont {M.~A.}\ \bibnamefont {Sentef}},\ }\href {\doibase
  10.1103/PhysRevResearch.1.023031} {\bibfield  {journal} {\bibinfo  {journal}
  {Phys. Rev. Research}\ }\textbf {\bibinfo {volume} {1}},\ \bibinfo {pages}
  {023031} (\bibinfo {year} {2019})}\BibitemShut {NoStop}%
\bibitem [{\citenamefont {He}\ \emph {et~al.}(2019)\citenamefont {He},
  \citenamefont {Addison}, \citenamefont {Jin}, \citenamefont {Mele},
  \citenamefont {Johnson},\ and\ \citenamefont {Zhen}}]{phth_He2019}%
  \BibitemOpen
  \bibfield  {author} {\bibinfo {author} {\bibfnamefont {L.}~\bibnamefont
  {He}}, \bibinfo {author} {\bibfnamefont {Z.}~\bibnamefont {Addison}},
  \bibinfo {author} {\bibfnamefont {J.}~\bibnamefont {Jin}}, \bibinfo {author}
  {\bibfnamefont {E.~J.}\ \bibnamefont {Mele}}, \bibinfo {author}
  {\bibfnamefont {S.~G.}\ \bibnamefont {Johnson}}, \ and\ \bibinfo {author}
  {\bibfnamefont {B.}~\bibnamefont {Zhen}},\ }\href {\doibase
  10.1038/s41467-019-12231-4} {\bibfield  {journal} {\bibinfo  {journal}
  {Nature Communications}\ }\textbf {\bibinfo {volume} {10}},\ \bibinfo {pages}
  {4194} (\bibinfo {year} {2019})}\BibitemShut {NoStop}%
\bibitem [{\citenamefont {Guglielmon}\ \emph {et~al.}(2018)\citenamefont
  {Guglielmon}, \citenamefont {Huang}, \citenamefont {Chen},\ and\
  \citenamefont {Rechtsman}}]{phexp_PhysRevA.97.031801}%
  \BibitemOpen
  \bibfield  {author} {\bibinfo {author} {\bibfnamefont {J.}~\bibnamefont
  {Guglielmon}}, \bibinfo {author} {\bibfnamefont {S.}~\bibnamefont {Huang}},
  \bibinfo {author} {\bibfnamefont {K.~P.}\ \bibnamefont {Chen}}, \ and\
  \bibinfo {author} {\bibfnamefont {M.~C.}\ \bibnamefont {Rechtsman}},\ }\href
  {\doibase 10.1103/PhysRevA.97.031801} {\bibfield  {journal} {\bibinfo
  {journal} {Phys. Rev. A}\ }\textbf {\bibinfo {volume} {97}},\ \bibinfo
  {pages} {031801} (\bibinfo {year} {2018})}\BibitemShut {NoStop}%
\bibitem [{\citenamefont {Banerjee}\ and\ \citenamefont
  {Narayan}(2020)}]{banerjee2020controlling}%
  \BibitemOpen
  \bibfield  {author} {\bibinfo {author} {\bibfnamefont {A.}~\bibnamefont
  {Banerjee}}\ and\ \bibinfo {author} {\bibfnamefont {A.}~\bibnamefont
  {Narayan}},\ }\href@noop {} {\bibfield  {journal} {\bibinfo  {journal} {arXiv
  preprint arXiv:2004.12606}\ } (\bibinfo {year} {2020})}\BibitemShut {NoStop}%
\bibitem [{\citenamefont {Firoz~Islam}\ and\ \citenamefont
  {Zyuzin}(2019)}]{3f_pi2_PhysRevB.100.165302}%
  \BibitemOpen
  \bibfield  {author} {\bibinfo {author} {\bibfnamefont {S.}~\bibnamefont
  {Firoz~Islam}}\ and\ \bibinfo {author} {\bibfnamefont {A.~A.}\ \bibnamefont
  {Zyuzin}},\ }\href {\doibase 10.1103/PhysRevB.100.165302} {\bibfield
  {journal} {\bibinfo  {journal} {Phys. Rev. B}\ }\textbf {\bibinfo {volume}
  {100}},\ \bibinfo {pages} {165302} (\bibinfo {year} {2019})}\BibitemShut
  {NoStop}%
\bibitem [{\citenamefont {{Wu}}\ \emph {et~al.}(2012)\citenamefont {{Wu}},
  \citenamefont {{Shi}},\ and\ \citenamefont {{Wei}}}]{tilt_gap_exp}%
  \BibitemOpen
  \bibfield  {author} {\bibinfo {author} {\bibfnamefont {J.-T.}\ \bibnamefont
  {{Wu}}}, \bibinfo {author} {\bibfnamefont {X.-H.}\ \bibnamefont {{Shi}}}, \
  and\ \bibinfo {author} {\bibfnamefont {Y.-J.}\ \bibnamefont {{Wei}}},\ }\href
  {\doibase 10.1007/s10409-012-0164-x} {\bibfield  {journal} {\bibinfo
  {journal} {Acta Mechanica Sinica}\ }\textbf {\bibinfo {volume} {28}},\
  \bibinfo {pages} {1539} (\bibinfo {year} {2012})},\ \Eprint
  {http://arxiv.org/abs/1210.7643} {arXiv:1210.7643 [cond-mat.mtrl-sci]}
  \BibitemShut {NoStop}%
\bibitem [{\citenamefont {Chan}\ \emph {et~al.}(2017)\citenamefont {Chan},
  \citenamefont {Lindner}, \citenamefont {Refael},\ and\ \citenamefont
  {Lee}}]{photocurrent_Weyl_PhysRevB.95.041104}%
  \BibitemOpen
  \bibfield  {author} {\bibinfo {author} {\bibfnamefont {C.-K.}\ \bibnamefont
  {Chan}}, \bibinfo {author} {\bibfnamefont {N.~H.}\ \bibnamefont {Lindner}},
  \bibinfo {author} {\bibfnamefont {G.}~\bibnamefont {Refael}}, \ and\ \bibinfo
  {author} {\bibfnamefont {P.~A.}\ \bibnamefont {Lee}},\ }\href {\doibase
  10.1103/PhysRevB.95.041104} {\bibfield  {journal} {\bibinfo  {journal} {Phys.
  Rev. B}\ }\textbf {\bibinfo {volume} {95}},\ \bibinfo {pages} {041104}
  (\bibinfo {year} {2017})}\BibitemShut {NoStop}%
\bibitem [{\citenamefont {Bruus}\ and\ \citenamefont
  {Flensberg}(2004)}]{bruus2004many}%
  \BibitemOpen
  \bibfield  {author} {\bibinfo {author} {\bibfnamefont {H.}~\bibnamefont
  {Bruus}}\ and\ \bibinfo {author} {\bibfnamefont {K.}~\bibnamefont
  {Flensberg}},\ }\href {https://books.google.co.in/books?id=zeaMBAAAQBAJ}
  {\emph {\bibinfo {title} {Many-Body Quantum Theory in Condensed Matter
  Physics: An Introduction}}},\ Oxford Graduate Texts\ (\bibinfo  {publisher}
  {OUP Oxford},\ \bibinfo {year} {2004})\BibitemShut {NoStop}%
\bibitem [{\citenamefont {Harper}\ \emph {et~al.}(2020)\citenamefont {Harper},
  \citenamefont {Roy}, \citenamefont {Rudner},\ and\ \citenamefont
  {Sondhi}}]{AnnRevCondMatt}%
  \BibitemOpen
  \bibfield  {author} {\bibinfo {author} {\bibfnamefont {F.}~\bibnamefont
  {Harper}}, \bibinfo {author} {\bibfnamefont {R.}~\bibnamefont {Roy}},
  \bibinfo {author} {\bibfnamefont {M.~S.}\ \bibnamefont {Rudner}}, \ and\
  \bibinfo {author} {\bibfnamefont {S.}~\bibnamefont {Sondhi}},\ }\href
  {\doibase 10.1146/annurev-conmatphys-031218-013721} {\bibfield  {journal}
  {\bibinfo  {journal} {Annual Review of Condensed Matter Physics}\ }\textbf
  {\bibinfo {volume} {11}},\ \bibinfo {pages} {345} (\bibinfo {year}
  {2020})}\BibitemShut {NoStop}%
\bibitem [{\citenamefont {Kuwahara}\ \emph {et~al.}(2016)\citenamefont
  {Kuwahara}, \citenamefont {Mori},\ and\ \citenamefont
  {Saito}}]{prethermal1Kuwahara_2016}%
  \BibitemOpen
  \bibfield  {author} {\bibinfo {author} {\bibfnamefont {T.}~\bibnamefont
  {Kuwahara}}, \bibinfo {author} {\bibfnamefont {T.}~\bibnamefont {Mori}}, \
  and\ \bibinfo {author} {\bibfnamefont {K.}~\bibnamefont {Saito}},\ }\href
  {\doibase 10.1016/j.aop.2016.01.012} {\bibfield  {journal} {\bibinfo
  {journal} {Annals of Physics}\ }\textbf {\bibinfo {volume} {367}},\ \bibinfo
  {pages} {96–124} (\bibinfo {year} {2016})}\BibitemShut {NoStop}%
\bibitem [{\citenamefont {Weidinger}\ and\ \citenamefont
  {Knap}(2017)}]{prethermal2Weidinger2017}%
  \BibitemOpen
  \bibfield  {author} {\bibinfo {author} {\bibfnamefont {S.~A.}\ \bibnamefont
  {Weidinger}}\ and\ \bibinfo {author} {\bibfnamefont {M.}~\bibnamefont
  {Knap}},\ }\href {\doibase 10.1038/srep45382} {\bibfield  {journal} {\bibinfo
   {journal} {Scientific Reports}\ }\textbf {\bibinfo {volume} {7}},\ \bibinfo
  {pages} {45382} (\bibinfo {year} {2017})}\BibitemShut {NoStop}%
\bibitem [{\citenamefont {Abanin}\ \emph {et~al.}(2017)\citenamefont {Abanin},
  \citenamefont {De~Roeck}, \citenamefont {Ho},\ and\ \citenamefont
  {Huveneers}}]{prethermal3Abanin2017}%
  \BibitemOpen
  \bibfield  {author} {\bibinfo {author} {\bibfnamefont {D.}~\bibnamefont
  {Abanin}}, \bibinfo {author} {\bibfnamefont {W.}~\bibnamefont {De~Roeck}},
  \bibinfo {author} {\bibfnamefont {W.~W.}\ \bibnamefont {Ho}}, \ and\ \bibinfo
  {author} {\bibfnamefont {F.}~\bibnamefont {Huveneers}},\ }\href {\doibase
  10.1007/s00220-017-2930-x} {\bibfield  {journal} {\bibinfo  {journal}
  {Communications in Mathematical Physics}\ }\textbf {\bibinfo {volume}
  {354}},\ \bibinfo {pages} {809} (\bibinfo {year} {2017})}\BibitemShut
  {NoStop}%
\bibitem [{\citenamefont {Iadecola}\ \emph {et~al.}(2015)\citenamefont
  {Iadecola}, \citenamefont {Neupert},\ and\ \citenamefont
  {Chamon}}]{heating1PhysRevB.91.235133}%
  \BibitemOpen
  \bibfield  {author} {\bibinfo {author} {\bibfnamefont {T.}~\bibnamefont
  {Iadecola}}, \bibinfo {author} {\bibfnamefont {T.}~\bibnamefont {Neupert}}, \
  and\ \bibinfo {author} {\bibfnamefont {C.}~\bibnamefont {Chamon}},\ }\href
  {\doibase 10.1103/PhysRevB.91.235133} {\bibfield  {journal} {\bibinfo
  {journal} {Phys. Rev. B}\ }\textbf {\bibinfo {volume} {91}},\ \bibinfo
  {pages} {235133} (\bibinfo {year} {2015})}\BibitemShut {NoStop}%
\bibitem [{\citenamefont {Dehghani}\ \emph {et~al.}(2015)\citenamefont
  {Dehghani}, \citenamefont {Oka},\ and\ \citenamefont
  {Mitra}}]{heating2PhysRevB.91.155422}%
  \BibitemOpen
  \bibfield  {author} {\bibinfo {author} {\bibfnamefont {H.}~\bibnamefont
  {Dehghani}}, \bibinfo {author} {\bibfnamefont {T.}~\bibnamefont {Oka}}, \
  and\ \bibinfo {author} {\bibfnamefont {A.}~\bibnamefont {Mitra}},\ }\href
  {\doibase 10.1103/PhysRevB.91.155422} {\bibfield  {journal} {\bibinfo
  {journal} {Phys. Rev. B}\ }\textbf {\bibinfo {volume} {91}},\ \bibinfo
  {pages} {155422} (\bibinfo {year} {2015})}\BibitemShut {NoStop}%
\bibitem [{\citenamefont {Seetharam}\ \emph {et~al.}(2015)\citenamefont
  {Seetharam}, \citenamefont {Bardyn}, \citenamefont {Lindner}, \citenamefont
  {Rudner},\ and\ \citenamefont {Refael}}]{heating3PhysRevX.5.041050}%
  \BibitemOpen
  \bibfield  {author} {\bibinfo {author} {\bibfnamefont {K.~I.}\ \bibnamefont
  {Seetharam}}, \bibinfo {author} {\bibfnamefont {C.-E.}\ \bibnamefont
  {Bardyn}}, \bibinfo {author} {\bibfnamefont {N.~H.}\ \bibnamefont {Lindner}},
  \bibinfo {author} {\bibfnamefont {M.~S.}\ \bibnamefont {Rudner}}, \ and\
  \bibinfo {author} {\bibfnamefont {G.}~\bibnamefont {Refael}},\ }\href
  {\doibase 10.1103/PhysRevX.5.041050} {\bibfield  {journal} {\bibinfo
  {journal} {Phys. Rev. X}\ }\textbf {\bibinfo {volume} {5}},\ \bibinfo {pages}
  {041050} (\bibinfo {year} {2015})}\BibitemShut {NoStop}%
\bibitem [{\citenamefont {Seetharam}\ \emph {et~al.}(2018)\citenamefont
  {Seetharam}, \citenamefont {Titum}, \citenamefont {Kolodrubetz},\ and\
  \citenamefont {Refael}}]{heating4PhysRevB.97.014311}%
  \BibitemOpen
  \bibfield  {author} {\bibinfo {author} {\bibfnamefont {K.}~\bibnamefont
  {Seetharam}}, \bibinfo {author} {\bibfnamefont {P.}~\bibnamefont {Titum}},
  \bibinfo {author} {\bibfnamefont {M.}~\bibnamefont {Kolodrubetz}}, \ and\
  \bibinfo {author} {\bibfnamefont {G.}~\bibnamefont {Refael}},\ }\href
  {\doibase 10.1103/PhysRevB.97.014311} {\bibfield  {journal} {\bibinfo
  {journal} {Phys. Rev. B}\ }\textbf {\bibinfo {volume} {97}},\ \bibinfo
  {pages} {014311} (\bibinfo {year} {2018})}\BibitemShut {NoStop}%
\bibitem [{\citenamefont {Nandy}\ \emph {et~al.}(2019)\citenamefont {Nandy},
  \citenamefont {Manna}, \citenamefont {C\ifmmode \u{a}\else
  \u{a}\fi{}lug\ifmmode~\u{a}\else \u{a}\fi{}ru},\ and\ \citenamefont
  {Roy}}]{tb_PhysRevB.100.235201}%
  \BibitemOpen
  \bibfield  {author} {\bibinfo {author} {\bibfnamefont {S.}~\bibnamefont
  {Nandy}}, \bibinfo {author} {\bibfnamefont {S.}~\bibnamefont {Manna}},
  \bibinfo {author} {\bibfnamefont {D.}~\bibnamefont {C\ifmmode \u{a}\else
  \u{a}\fi{}lug\ifmmode~\u{a}\else \u{a}\fi{}ru}}, \ and\ \bibinfo {author}
  {\bibfnamefont {B.}~\bibnamefont {Roy}},\ }\href {\doibase
  10.1103/PhysRevB.100.235201} {\bibfield  {journal} {\bibinfo  {journal}
  {Phys. Rev. B}\ }\textbf {\bibinfo {volume} {100}},\ \bibinfo {pages}
  {235201} (\bibinfo {year} {2019})}\BibitemShut {NoStop}%
\bibitem [{\citenamefont {Bernevig}\ and\ \citenamefont
  {Hughes}(2013)}]{bernevig2013topological}%
  \BibitemOpen
  \bibfield  {author} {\bibinfo {author} {\bibfnamefont {B.}~\bibnamefont
  {Bernevig}}\ and\ \bibinfo {author} {\bibfnamefont {T.}~\bibnamefont
  {Hughes}},\ }\href {https://books.google.co.in/books?id=\_7r\_UqFN0IEC}
  {\emph {\bibinfo {title} {Topological Insulators and Topological
  Superconductors}}}\ (\bibinfo  {publisher} {Princeton University Press},\
  \bibinfo {year} {2013})\BibitemShut {NoStop}%
\end{thebibliography}
%

\end{document}